\documentclass[twocolumn,trackchanges]{aastex63}
\usepackage{amsmath}
\usepackage{amssymb}
\usepackage{braket}
\usepackage{amssymb}
\usepackage{color}
\usepackage{xcolor}
\usepackage{graphicx}
\usepackage{latexsym}
\usepackage{listings}
\usepackage{mathrsfs}
\usepackage{aas_macros}
\usepackage{letltxmacro}
\usepackage[normalem]{ulem}

\usepackage{subfigure}
\usepackage{verbatim}
\usepackage{tabularx}
\usepackage{ragged2e}
\usepackage{multirow}
\usepackage{amsbsy}

\usepackage{numprint}
\usepackage{makecell}
\usepackage[utf8]{inputenc}

\usepackage[shortlabels]{enumitem}

\newcommand{\R}{\mathbb{R}}
\newcommand{\N}{\mathbb{N}} % Integers
\newcommand{\paren}[1]{\left( #1 \right)}
\newcommand{\sqrbrace}[1]{\left[ #1 \right]}

\newcommand{\dwmean}[1]{\left\langle #1 \right\rangle_{\rho,r,\theta,\phi}}
\newcommand{\dwstd}[1]{\text{std}\left( #1 \right)_{\rho,r,\theta,\phi}}

\renewcommand{\braket}[1]{\left\langle #1 \right\rangle}

\newcommand{\V}{\mathcal{V}}

\newcommand{\F}{\mathcal{F}}

\newcommand{\M}{\mathcal{M}}
\newcommand{\I}{\mathcal{I}}

\newcommand{\B}{\text{B}}

\newcommand{\nubhlight}{$\nu\texttt{bhlight}$${}$}

\newcommand{\detg}{\sqrt{-g}}

\newcommand{\eepsilon}{\epsilon} % energy

\newcommand{\jnuf}{j_{\eepsilon,f}}
\newcommand{\etanuf}{\eta_{\eepsilon,f}}
\newcommand{\Inuf}{I_{\eepsilon,f}}
\newcommand{\chinuf}{\chi_{\eepsilon,f}}
\newcommand{\sigmanuf}{\sigma_{\eepsilon,f}}
\newcommand{\alphanuf}{\alpha_{\eepsilon,f}}

\npdecimalsign{.}
\npthousandsep{,}

\newcolumntype{Y}{>{\RaggedRight\arraybackslash}X}

\graphicspath{{./figures/}}

\sloppypar

\submitjournal{ApJ}

\shorttitle{GRRMHD for Collapsars}
\shortauthors{Miller et al.}

\begin{document}

\title{Full Transport General Relativistic Radiation
  Magnetohydrodynamics for Nucleosynthesis in Collapsars}

\author{Jonah M. Miller}
\email{jonahm@lanl.gov}
\affiliation{Computational Physics and Methods, Los Alamos National Laboratory, Los Alamos, NM, USA}
\affiliation{Center for Theoretical Astrophysics, Los Alamos National Laboratory, Los Alamos, NM, USA}
\affiliation{Center for Nonlinear Studies, Los Alamos National Laboratory, Los Alamos, NM, USA}

\author{Trevor M. Sprouse}
\affiliation{Los Alamos Center for Space and Earth Science Student Fellow}
\affiliation{Department of Physics, University of Notre Dame, Notre
  Dame, IN, USA}
\affiliation{Theoretical Divison, Los Alamos National Laboratory, Los Alamos, NM 87545, USA}

\author{Christopher L. Fryer}
\affiliation{Computational Physics and Methods, Los Alamos National Laboratory, Los Alamos, NM 87545, USA}
\affiliation{Center for Theoretical Astrophysics, Los Alamos National Laboratory, Los Alamos, NM 87545, USA}

\author{Benjamin R. Ryan}
\affiliation{Computational Physics and Methods, Los Alamos National Laboratory, Los Alamos, NM, USA}
\affiliation{Center for Theoretical Astrophysics, Los Alamos National Laboratory, Los Alamos, NM, USA}

\author{Joshua C. Dolence}
\affiliation{Computational Physics and Methods, Los Alamos National Laboratory, Los Alamos, NM, USA}
\affiliation{Center for Theoretical Astrophysics, Los Alamos National Laboratory, Los Alamos, NM, USA}

\author{Matthew R. Mumpower}
\affiliation{Theoretical Divison, Los Alamos National Laboratory, Los Alamos, NM 87545, USA}
\affiliation{Center for Theoretical Astrophysics, Los Alamos National Laboratory, Los Alamos, NM 87545, USA}

\author{Rebecca Surman}
\affiliation{Department of Physics, University of Notre Dame, Notre
  Dame, IN, USA}

\begin{abstract}
  We model a compact black hole-accretion disk system in the collapsar
  scenario with full transport, frequency dependent, general
  relativistic radiation magnetohydrodynamics. We examine whether or
  not winds from a collapsar disk can undergo rapid neutron capture
  (r-process) nucleosynthesis and significantly contribute to solar
  r-process abundances. We find the inclusion of accurate transport
  has significant effects on outflows, raising the electron fraction
  above $Y_{\rm e} \sim 0.3$ and preventing third peak r-process material
  from being synthesized. We analyze the time-evolution of neutrino
  processes and electron fraction in the disk and present a simple
  one-dimensional model for the vertical structure that emerges. We
  compare our simulation to semi-analytic expectations and argue that
  accurate neutrino transport and realistic initial and boundary
  conditions are required to capture the dynamics and nucleosynthetic
  outcome of a collapsar.
\end{abstract}

\section{Introduction}
\label{sec:intro}

When a massive, rapidly rotating star collapses, it mail fail to
explode. Post-bounce, the proto-neutron star collapses and forms a
black hole.  In this scenario, stellar material eventually
circularizes and accretes onto the central black
hole. \citet{Woosley1993} coined this a ``failed'' supernova, with
``failed'' in quotes, since an accretion-driven jet may indeed cause
an explosion.  \citet{MacFadyen_1999} coined this the collapsar
scenario, and this system a collapsar. These events are commonly
invoked as the sources of long gamma ray bursts (GRBs), and
observational evidence is consistent with this hypothesis
\citep{WoosleySNGRbConnection,GhirlandaGRBMechanisms,GRBSNConnection}.

The dynamics of stellar collapse and the formation of a GRB engine has
thus been studied extensively, see
\citet{Woosley1993,MacFadyen_1999,MacFayden2001,Proga2003,Heger2003B,Mizuno2004GRMHDCollapsar,Fujimoto2006,Nagataki2007Collapsar,Rockefeller2006Collapsar3D,UzdenskyMacFayden2007,Morsony2007,BucciantiniQuataert2008,LazzatiGRB2008,Kumar2008Fallback,Nagakura2009JetBreakout,Taylor2011CollapsarGADGET,OttGRCollapsar,Lindner2012,LopezAMR2013,Batta2014Cooling,NDAFReview}
and references therein. Recently, attention has been devoted to the
related case where a rapidly rotating star collapses to a protoneutron
star and black hole formation is either delayed or does not happen at
all
\citep{ThompsonMagnetarSpindown2004,MetzgerThompsonQuataert2008MagnetarOutflow,Winteler2012MHDSN,Moesta2014SN3D,Moesta2018GRMDH,Halevi2018GRMHD}.

\begin{figure}[t!]
  \centering
  \includegraphics[width=\linewidth]{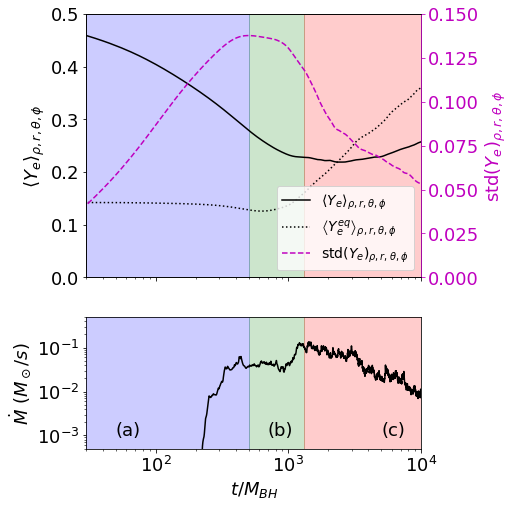}
  \caption{Density-weighted mean and standard deviation of electron
    fraction $Y_{\rm e}$ (top) and accretion rate $\dot{M}$ (bottom) as a
    function of time. The dotted black line shows the density-weighted
    mean \textit{equilibrium} value of $Y_{\rm e}$ as opposed to the one
    realized in the simulation. The evolution of the disk can be
    roughly broken into three phases, (a), (b), and (c), described in
    the text.}
  \label{fig:time:evolution}
\end{figure}

\citet{MacFadyen_1999} realized that the dynamics of collapsar disks
are similar to other neutron-rich compact accretion flows such as
those formed by a binary neutron star merger. This implies that
collapsar disks may be a proposed site of rapid neutron capture
(r-process) nucleosynthesis, the mechanism by which the heaviest
elements in our universe are formed
\citep{HowardRareNuclei,LattimerCBC1,LattimerCBC2,Blinnikov,KohriDisks,CoteRProcess}.\footnote{For
  recent review of the r-process, see \citet{CowanReview}.}
Nucleosynthesis in rapidly rotating core collapse---with and without
black hole formation---has been explored by several groups
\citep{PophamNDAF,dimatteo02,surman04,mclaughlin05,surman06,ThompsonMagnetarSpindown2004,Rockefeller2008NucleosynthesisFromCollapsars,MetzgerThompsonQuataert2008MagnetarOutflow,Winteler2012MHDSN,Moesta2014SN3D,Moesta2018GRMDH,Halevi2018GRMHD}.

At low entropies, electron fraction becomes the deciding factor in the
r-process. For r-process elements to be synthesized, the central
engine must produce outflows with low electron fraction $Y_{\rm e}$. A
robust r-process typically requires $Y_{\rm e} \lesssim 0.25$. Early
semi-analytic work found that collapsar outflows are insufficiently
neutron-rich
\citep{PophamNDAF,dimatteo02,surman04,mclaughlin05,surman06}. In the
magnetar case, where no black hole formation occurs, three-dimensional
simulations show that non-axisymmetric effects can make it difficult
to eject a sufficient amount of low $Y_{\rm e}$ material. Thus whether or
not a magnetar can eject heavy elements depends on factors that
control the symmetry of the problem, such as magnetic field strength
and how quickly the jet develops
\citep{Moesta2014SN3D,Moesta2018GRMDH,Halevi2018GRMHD}.

One proposed mechanism for producing massive, neutron-rich outflows is
that material may be entrained in a low-density relativistic jet
\citep{Fujimoto2007,Ono2012,Nakamura2015,Soker2017,Kayakawa2018}. One
promising aspect of this approach is that material entrained in the
jet may have high entropy, which means that it may undergo rapid
neutron capture even with higher $Y_{\rm e}$. Most of these works assume
axisymmetry, which means they do not properly account for the
non-axisymmetric kink instability \citep{Moesta2014SN3D} and suffer
from the anti-dynamo theorem
\citep{CowlingDynamo1,CowlingDynamo2}. Another issue is that if a jet
is loaded with too much material, it cannot reach large Lorentz
factors, meaning there is a tension between producing a robust jet
and producing a sufficient amount of r-process material. It remains to
be seen whether this mechanism holds up for realistic
three-dimensional models and whether or not it can provide a
meaningful contribution to abundances of r-process elements in the
universe.

Recently \citet{SiegelCollapsar} argued that collapsar fallback and
subsequent accretion onto the central black hole can be approximately
modeled by a magnetohydrodynamically driven accretion disk. They
performed a suite of three-dimensional magnetohydrodynamic
simulations, each corresponding to a different accretion rate, and
thus a different phase of the core-collapse fallback. They find that
the outflow from their simulation with the highest accretion rate is
neutron-rich and they use this result to argue that collapsars are a
primary source of r-process elements in the universe. In addition to
the nucleosynthetic implications, \citet{SiegelCollapsar} make an
observable prediction about long GRBs. Assuming a long GRB is driven
by a collapsar, the radioactive decay of r-process elements from the
outflow implies an infra-red excess in the afterglow of
such an event.

\citet{SiegelCollapsar} modeled neutrino radiation with a leakage
scheme first described in \citet{SiegelMetzger3DBNS} and based on a
long lineage
\citep{Bruenn1985nu,Ruffert1997DiskWind,GaleazziLeakage,RadiceBNS2016}. However,
neutrino transport can have significant effects on the electron
fraction and nucleosynthesis in compact accretion flows
\citep{MillerGW170817}. We therefore wish to see how improved
transport effects the collapsar scenario. We model the highest
accretion rate and thus densest, highest temperature, lowest electron
fraction and most nucleosynthetically optimistic disk from
\citet{SiegelCollapsar} with full frequency dependent general
relativistic neutrino radiation magnetohydrodynamics. We then perform
r-process nucleosynthesis calculations on the resulting outflow in
post-processing.

We find that neutrino transport has significant effects on the disk
outflow. In particular, although rapid neutron capture occurs, $Y_{\rm e}$
is not low enough in the outflow to produce third-peak r-process
material. We also use our model to explore the hypothesis that a
compact accretion disk is a sufficiently descriptive surrogate for a
full collapsar. Although we are unable to make strong claims on the
validity of using a single disk as a proxy for a collapsar, we argue
that models with better initial and boundary conditions will continue
to lack 3rd-peak r-process elements. However, further work is required
to more deeply understand the system as a whole.

In section \ref{sec:system}, we describe the physical system we
simulate. In section \ref{sec:methods}, we describe our numerical
method and discuss resolution requirements. In section
\ref{sec:results}, we present results from our simulation, including
steady-state disk properties, outflow statistics, and nucleosynthetic
yield. In section \ref{sec:systematics}, we examine systematic
effects in our simulation. We discuss the importance of full neutrino
transport and neutrino absorption in achieving our steady-state disk
and outflow properties and we comment on the influence of the initial
and boundary conditions and discuss the prospect of outflow material
escaping the star. Finally, in section \ref{sec:conclusions}, we
summarize our results and discuss some implications of our work.

\begin{figure}[t!]
  \centering
  \includegraphics[width=\linewidth]{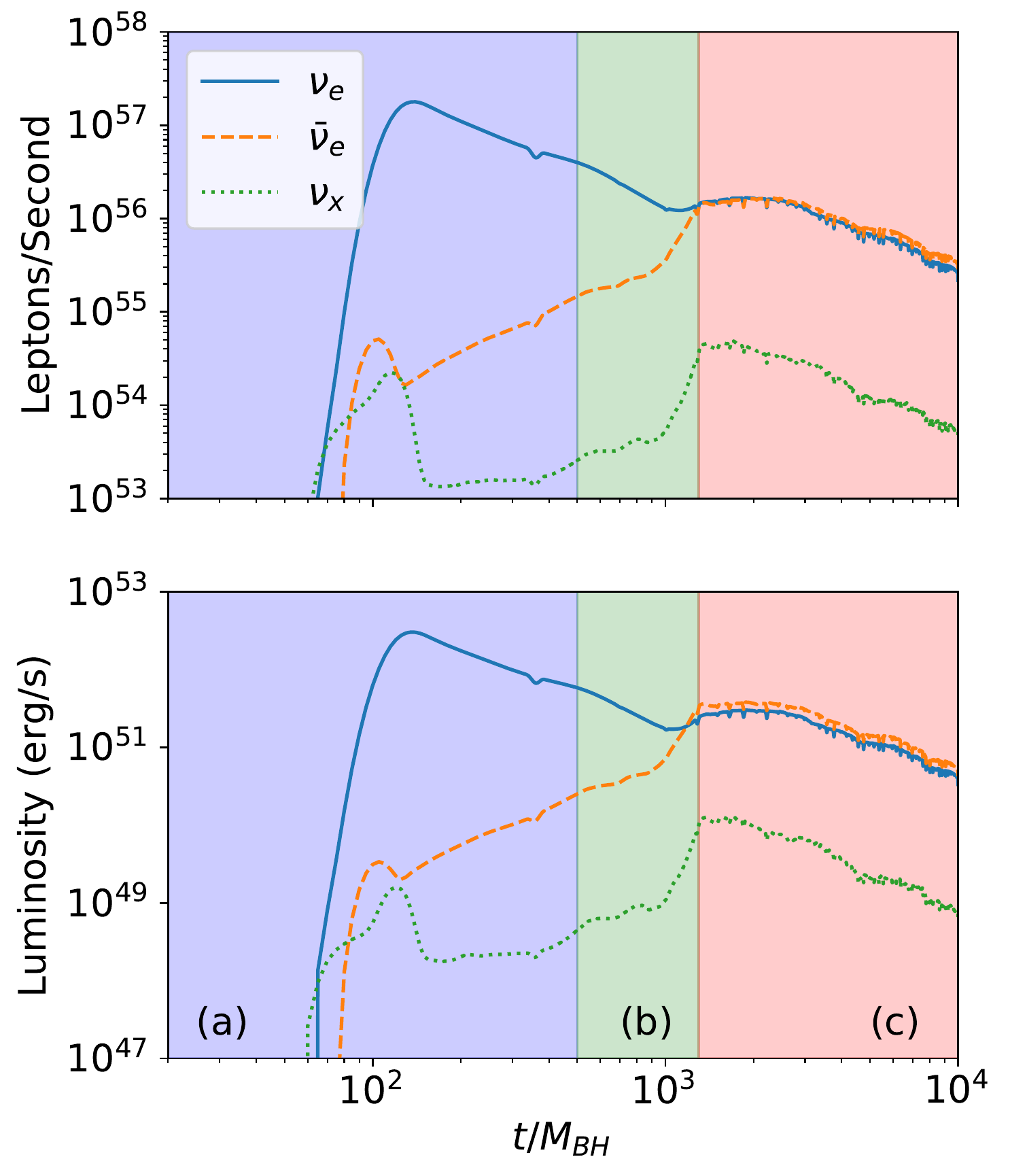}
  \caption{Luminosity (bottom) and particles per second (top) for
    electron neutrinos, electron antineutrinos, and heavy neutrinos
    measured at infinity as functions of time. The phases identified
    in Figure \ref{fig:time:evolution} are shown. The luminosity for
    heavy neutrinos is universally orders of magnitude lower than for
    the electron neutrinos and their antiparticles. At early times,
    the luminosity for electron neutrinos is much higher than for
    electron anti-neutrinos, consistent with rapid deleptonization. At
    late times, electron neutrinos and their antiparticles are roughly
    in balance, with a slight excess in anti-neutrinos, consistent
    with a slow releptonization process.}
  \label{fig:nu:luminosity}
\end{figure}

\begin{table*}[t!]
  \centering
  \begin{tabular}{l | l | l | l}
    \hline
    \textbf{Type}&\textbf{Processes}&\textbf{Charged/Neutral}&\textbf{Corrections/Approximations}\\
    \hline
    % \textbf{Emission}&&\\
    \hline
    Electron Capture on Protons & 
                             \begin{tabular}{@{}l@{}}
                               $\nu_e + n \leftrightarrow e^- + p$\\
                               $\nu_\mu + n \leftrightarrow \mu^- + p$
                             \end{tabular}
                 & Charged &
                             \begin{tabular}{@{}l@{}}
                               Blocking/Stimulated
                               Abs.\\
                               Weak
                               Magnetism\\
                               Recoil
                             \end{tabular}\\
    \hline
    Positron Capture on Neutrons & 
                            \begin{tabular}{@{}l@{}}
                              $\bar{\nu}_e + p \leftrightarrow e^+ +
                              n$\\
                              $\bar{\nu}_\mu + p \leftrightarrow \mu^+
                              + n$\\
                            \end{tabular}
                 & Charged &
                             \begin{tabular}{@{}l@{}}
                               Blocking/Stimulated
                               Abs.\\
                               Weak
                               Magnetism\\
                               Recoil
                             \end{tabular}\\
    \hline
    Abs./Emis. on Ions & $\nu_eA \leftrightarrow A'e^-$ & Charged &
                                                                    \begin{tabular}{@{}l@{}}
                                                                      Blocking/Stimulated Abs.\\
                                                                      Recoil
                                                                    \end{tabular}\\
    \hline
    Electron Capture on Ions & $e^- + A \leftrightarrow A' + \nu_e$ &
                                                                      Charged &
                                                                                \begin{tabular}{@{}l@{}}
                                                                                  Blocking/Stimulated Abs.\\
                                                                                  Recoil
                                                                                \end{tabular}\\
                                                             
    \hline
    $e^+-e^-$ Annihilation & $e^+e^- \leftrightarrow \nu_i\bar{\nu}_i$&
                                                                    Charged
                                                                    + Neutral &
                                                                    \begin{tabular}{@{}l@{}}
                                                                      single-$\nu$
                                                                      Blocking\\
                                                                      Recoil
                                                                    \end{tabular}\\
    \hline
    $n_i$-$n_i$ Brehmsstrahlung & $n_i^1 + n_i^2 \to n_i^3 +
                                      n_i^4 + \nu_i\bar{\nu}_i$ 
                                    & Neutral & 
                                                \begin{tabular}{@{}l@{}}
                                                  single-$\nu$ Blocking\\
                                                  Recoil
                                                \end{tabular}\\
    \hline
    
  \end{tabular}
  \caption{Emission and Absorption Processes used in \nubhlight.}
  \tablecomments{The symbols in the processes are as follows: $n$ is a
    neutron, $p$ a proton, $e^-$ an electron, $e^+$ a proton, $\mu^-$
    a muon, $\mu^+$ an antimuon, and $n_i$ an arbitrary
    nucleon. $\nu_i$ is an arbitrary neutrino. $\nu_e$ is an electron
    neutrino, and $\bar{\nu}_e$ is an electron antineutrino.
    %Muon processes are only relevant for high temperatures, above 100 MeV. 
    % Emission and
    % absorption on protons are the most dominant processes for the
    % densities and temperatures of a post-merger disk. 
    We describe the corrections and approximations used below, as
    tabulated in \citet{fornax} and provided to us in
    \citet{BurrowsCorresp}. Blocking and stimulated absorption are
    related to the Fermi-Dirac nature of neutrinos. Weak magnetism is
    related to the extended quark structure of nucleons. Recoil is the
    kinematic recoil. Single-$\nu$ blocking is a Kirkhoff's law based
    approximation of blocking that becomes exact for processes that
    involve only a single neutrino. The details of these interactions
    are summarized in \citet{BurrowsNeutrinos}. This table was first
    presented in its current form in \citet{nubhlight}.}
  \label{tab:emis:abs}
\end{table*}

\section{The Model}
\label{sec:system}

We solve the equations of general relativistic ideal
magnetohydrodynamics (MHD)\footnote{Unless otherwise noted, we assume
  Greek indices range over spacetime, from 0 to 3, and Latin indices
  range over space, from 1 to 3.}
\begin{widetext}
\begin{eqnarray}
  \label{eq:particle:cons}
  \partial_t \paren{\detg\rho u^t} + \partial_i\paren{\detg\rho u^i}
  &=& 0\\
  \label{eq:energy:cons}
  \partial_t\sqrbrace{\detg \paren{T^t_{\ \nu} + \rho u^t \delta^t_\nu}}
  + \partial_i\sqrbrace{\detg\paren{T^i_{\ \nu} + \rho u^i \delta^t_\nu}}
  &=& \detg \paren{T^\kappa_{\ \lambda} \Gamma^\lambda_{\nu\kappa} + G_\nu}\ \forall \nu = 0,1,2,3\\
  \label{eq:mhd:cons}
  \partial_t \paren{\detg B^i} + \partial_j \sqrbrace{\detg\paren{b^ju^i - b^i u^j}} &=& 0\ \forall i=1,2,3\\
  \label{eq:lepton:cons}
  \partial_t\paren{\detg\rho Y_{\rm e} u^t} + \partial_i\paren{\detg\rho Y_{\rm e}u^i}
  &=& \detg G_{Y_{\rm e}}
\end{eqnarray}
\end{widetext}
where the energy-momentum tensor $T^\mu_{\ \nu}$ is assumed to
be
\begin{equation}
  \label{eq:def:Tmunu}
  T^\mu_{\ \nu} = \paren{\rho + u + P + b^2}u^\mu u_\nu + \paren{P + \frac{1}{2} b^2} \delta^\mu_\nu - b^\mu b_\nu
\end{equation}
for metric $g_{\mu\nu}$, rest energy $\rho$, fluid four-velocity
$u^\mu$, internal energy density $u$, pressure $P$, and Christoffel
connection $\Gamma^\alpha_{\beta\gamma}$. (Here and in the remainder
of the text, unless otherwise specified, we set $G=c=1$.)

Equation \eqref{eq:particle:cons} represents conservation of baryon
number. Equation \eqref{eq:energy:cons} represents conservation of
energy-momentum, subject to the radiation four-force $G_\nu$. Equation
\eqref{eq:mhd:cons} describes the evolution of magnetic fields, where
\begin{equation}
  \label{eq:def:Bi}
  B^i = ^*F^{it}
\end{equation}
comprise the magnetic field components of the Maxwell tensor
$F_{\mu\nu}$ and $b^\mu$ is the magnetic field four-vector
\begin{equation}
  \label{eq:def:bmu}
  ^*F^{\mu\nu} = b^\mu u^\nu - b^\nu u^\mu.
\end{equation}
Equation \eqref{eq:lepton:cons} describes the conservation of
lepton number. $G_{ye}$ is a source term describing the rate at which
lepton density is transferred between the fluid and the radiation field.

We close our system with the SFHo equation of state, tabulated in the
Stellar Collapse format
\citep{stellarcollapsetables,stellarcollapseweb} and described in
\citet{SFHoEOS}. An equation of state relates the pressure $P$ and
specific internal energy $\varepsilon$ to the density $\rho$,
temperature $T$, and electron fraction $Y_{\rm e}$:
\begin{eqnarray}
  \label{eq:def:EOS}
  P &=& P(\rho, T, Y_{\rm e})\\
  \label{eq:def:EOS:e}
  \varepsilon &=& \varepsilon(\rho, T, Y_{\rm e}).
\end{eqnarray}
We evolve $\rho$, $u = \rho\varepsilon$, and $Y_{\rm e}$, but not $T$ or
$P$. So at a given time, we find $T$ by inverting equation
\eqref{eq:def:EOS:e} before plugging it into equation
\eqref{eq:def:EOS} to find $P$.

We approximate our neutrinos as massless such that they obey the
standard radiative transfer equation
\begin{equation}
  \label{eq:radiative:transfer}
  \frac{D}{d\lambda}\paren{\frac{h^3\Inuf}{\eepsilon^3}} = \paren{\frac{h^2\etanuf}{\eepsilon^2}} - \paren{\frac{\eepsilon \chinuf}{h}} \paren{\frac{h^3\Inuf}{\eepsilon^3}},
\end{equation}
where $D/d\lambda$ is the derivative along a neutrino trajectory in
phase space, $\Inuf$ is the specific intensity of the neutrino field of
flavor $f\in \{\nu_e,\bar{\nu}_e,\nu_x\}$,
\begin{equation}
  \label{eq:def:extinction:coeff}
  \chinuf = \alphanuf + \sigmanuf^a
\end{equation}
is the extinction coefficient that combines absorption coefficient
$\alphanuf$ and scattering extinction $\sigmanuf^a$ for scattering
interaction $a$ and
\begin{equation}
  \label{eq:def:emission:coeff}
  \etanuf = \jnuf + \etanuf^s(\Inuf)
\end{equation}
is the emissivity combining fluid emissivity $\jnuf$ and emission due
to scattering from $\etanuf^s$. Here $h$ is Planck's constant,
$\eepsilon$ is the energy of a neutrino with wavevector $k^\mu$ as
measured by an observer traveling along a timelike Killing vector
$\eta^\mu$.

Neutrinos can interact with matter via emission, absorption, or
scattering. The latter does not change electron fraction $Y_{\rm e}$, while
the former two do. For emission and absorption, we use the charged and
neutral current interactions as tabulated in \citet{fornax} and
described in \citet{BurrowsNeutrinos}. We summarize these interactions
in Table \ref{tab:emis:abs}, which was first presented in
\citet{nubhlight}. We treat inelastic scattering off of electrons,
nucleons, and heavy nuclei. Our scattering implementation uses a
biasing technique to ensure all processes are well sampled, as
described in \citet{nubhlight}.

\begin{figure}[t!]
  \centering
  \includegraphics[width=\linewidth]{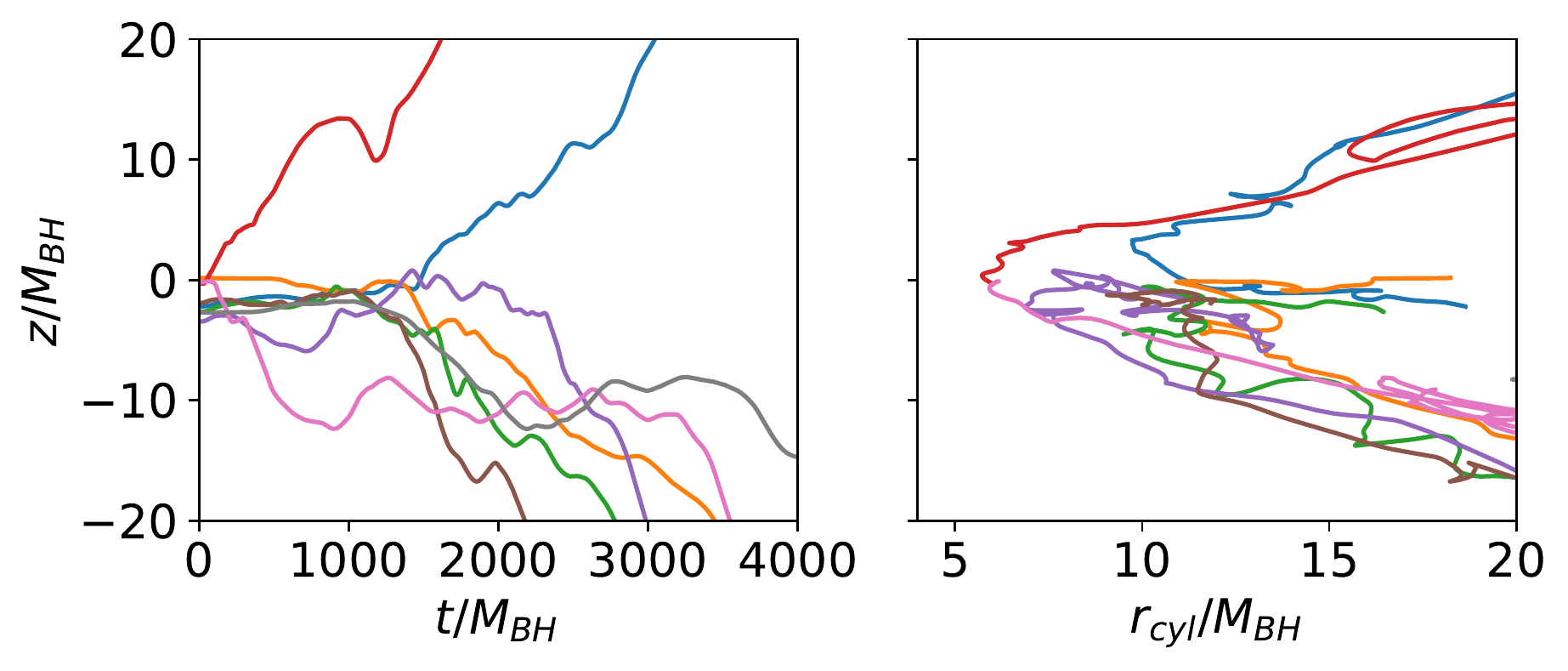}
  \caption{A random selection of traces that vary in latitude within
    the disk. We show height $z$ as a function of time (left) and
    cylindrical radius $r_{\rm cyl} = \sqrt{x^2 + y^2}$ (right). Notice
    that tracer paths are not linear. Rather, an individual fluid
    packet appears to random walk through space.}
  \label{fig:selected:traces:latitude}
\end{figure}

\section{Methods}
\label{sec:methods}

We simulate a disk of accretion rate
$\dot{M} \approx 10^{-1} M_\odot/s$ in a stationary Kerr black hole
spacetime \citep{KerrBH} for a black hole of mass $M_{\rm BH} = 3 M_\odot$
and dimensionless spin $a=0.8$, corresponding to the most
nucleosynthetically optimistic (and highest $\dot{M}$) case presented
in \citet{SiegelCollapsar}. To form the accretion disk, we begin with
a torus in hydrostatic equilibrium \citep{FishboneMoncrief} of
constant specific angular momentum, constant entropy of
$s = 8 k_b/$baryon, constant electron fraction $Y_{\rm e}=0.5$, and total
mass of $M_{\rm d} = 0.02 M_\odot$. These conditions imply our torus has an
inner radius of 5.5 $GM_{\rm BH}/c^2$ and a radius of peak pressure of
12.525 $GM_{\rm BH}/c^2$. Our torus starts with a single poloidal magnetic
field loop with a minimum ratio of gas to magnetic pressure $\beta$ of 100. As
the system evolves, the magneto-rotational instability
\citep[MRI,][]{BalbusHawley91} self-consistently drives the disk to a
turbulent state, which provides the turbulent viscosity necessary for
the disk to accrete \citep{ShakuraSunyaevAlpha}.

\begin{figure}[tb!]
  \centering
  \includegraphics[width=\linewidth]{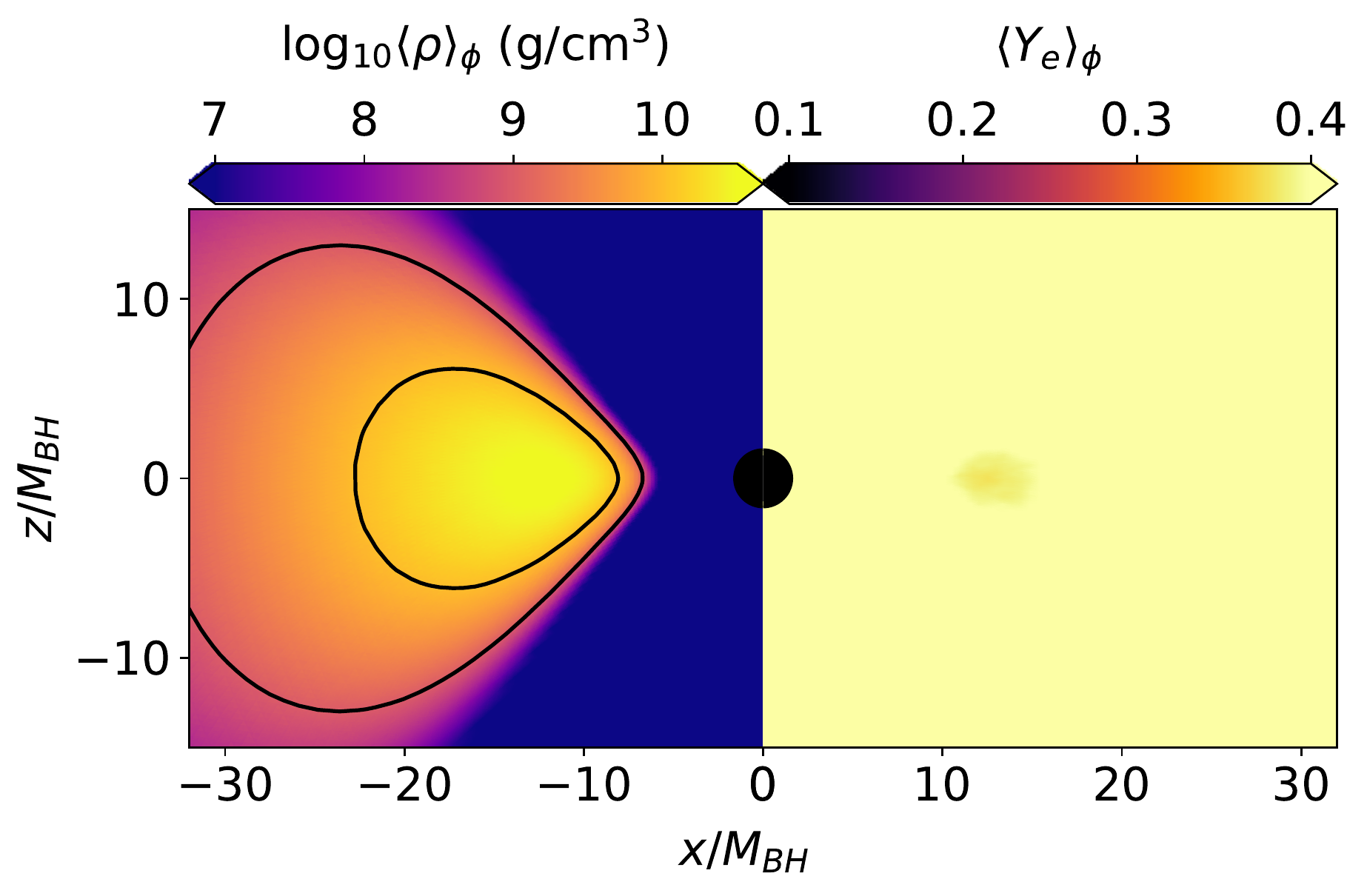}
  \caption{Density $\rho$ (left) and electron fraction $Y_{\rm e}$
    (right) near the initial time, at $t = 25 G M_{\rm BH}/c^3$ or
    $\approx 0.3$ ms. Contours are for $\rho=10^9$ and $10^{10}$
    g$/$cm$^3$. At this time, $Y_{\rm e}$ is almost universally 0.5
    and deleptonization is beginning. Quantities are averaged
    over azimuthal angle $\phi$.}
  \label{fig:rho:ye:t0}
\end{figure}

We use our code \nubhlight~\citep{nubhlight}, based on
\texttt{bhlight} \citep{bhlight}, which uses operator splitting to
couple GRMHD via finite volume methods with constrained transport
\citep{HARM} to neutrino transport via Monte Carlo methods
\citep{grmonty}. We use a radially logarithmic, quasi-spherical grid
in horizon penetrating coordinates, as first presented in \citet{HARM}
with $N_r\times N_\theta \times N_\phi = 192\times 168\times 66$ grid
points with approximately $3\times 10^7$ Monte Carlo packets. Although
our code is Eulerian, we track Lagrangian fluid packets with
approximately $1.5\times 10^6$ ``tracer particles,'' of which
approximately $5\times 10^5$ become gravitationally unbound. Following
\citet{BovardTracersBNS}, our tracer particles are initialized within
the disk so that they uniformly sample disk material by volume. For
more information on our code implementation and verification, see
\citet{nubhlight}. For a first application of \nubhlight${}$ in the
context of neutron star mergers, see \citet{MillerGW170817}.

\begin{figure}[tb!]
  \centering
  \includegraphics[width=\linewidth]{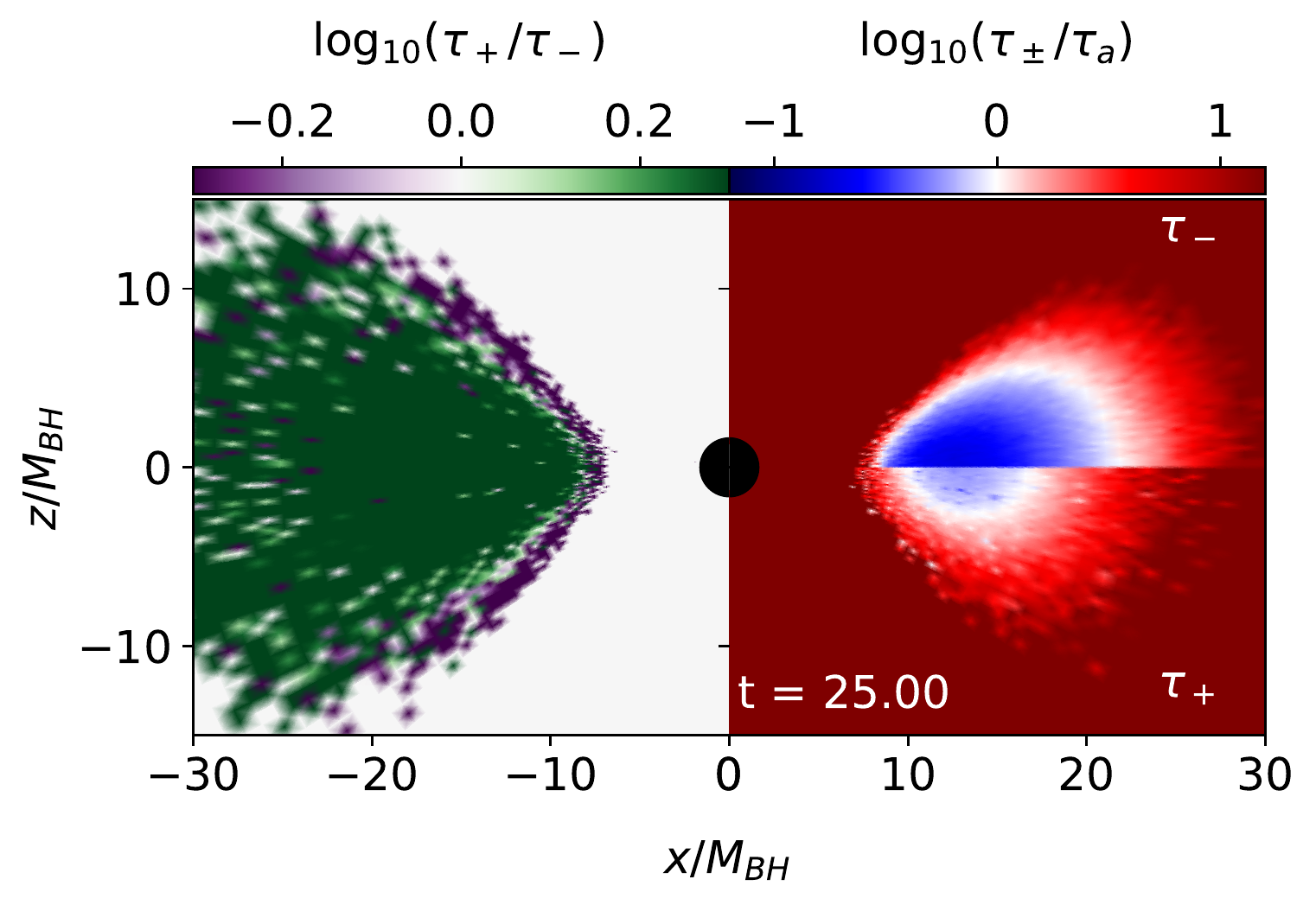}
  \caption{Comparison of relevant time scales for processes that can
    change $Y_{\rm e}$: an increase due to radiation $\tau_+$, a decrease
    due to radiation $\tau_-$ and change due to turbulent flow
    $\tau_a$. Plotted are: ratio of $\tau_+/\tau_-$ (left), ratio of
    $\tau_-/\tau_a$ (top right), and ratio of $\tau_+/\tau_a$ (bottom
    right), near the initial time, at $t = 25 G M_{\rm BH}/c^3$ or
    $\approx 0.3$ ms and averaged over azimuthal angle $\phi$.}
  \label{fig:time:scales:t0}
\end{figure}

We run our simulation for approximately $10^4 G M_{\rm BH}/c^3$, or 148
ms, which allows us to observe the disk in a quasistationary turbulent
state. In the collapsar paradigm, the disk is fed by circularized
fallback from the progenitor star as it undergoes gravitational
collapse. The initial phase of our simulation, where we relax an
equilibrium torus, comprises an unphysical transient, and we wish to
ignore material driven off the disk in this transient phase. We
therefore neglect outflow which reaches a surface of
$r = 250 G M_{\rm BH}/c^2$ within the first half of the simulation,
$t < 5\times 10^3 G M_{\rm BH}/c^3$. Note that this corresponds to
material ejected from the disk at much earlier times. We experimented
with the amount of time we neglect and found that it does not
significantly change the results presented below.

An accurate magnetohydrodynamic model of turbulent viscosity requires
capturing the MRI \citep{BalbusHawley91}. Following \citet{Sano2004},
we define a quality factor
\begin{equation}
  \label{eq:q:theta}
  Q^{(\theta)}_{\text{mri}} = \frac{2\pi b^{(\theta)}}{\Delta x^{(\theta)}\sqrt{w + b^2}\Omega},
\end{equation}
for the MRI to be the number of grid points per minimum unstable MRI
wavelength inside the disk. Here $b^{(\theta)}$ is the
$\theta$-component of the magnetic field four-vector,
$\Delta x^{(\theta)}$ is grid spacing in the $\theta$ direction, $w$
is the enthalpy of the fluid, $\Omega$ is the angular velocity of the
flow, and $b^2 = b^\mu b_\mu$ is total magnetic field strength. To
resolve the MRI, one needs at least ten grid points per smallest
unstable MRI wavelength \citep{Hawley2013}. Our measurements of our
disk satisfy this requirement, with
$Q^{(\theta)}_{\text{mri}} \geq 10$ within the disk for all times.

Our simulation is not only magnetohydrodynamic, but \textit{radiation}
magnetohydrodynamic. Therefore, it is important also to ensure we are
using a sufficient number of Monte Carlo packets to capture the
relevant interactions between the gas and radiation field. Following
\citet{MillerGW170817}, we define the Monte Carlo quality factor
\begin{equation}
  \label{eq:quality:factor}
  Q_{\text{rad}} = \min_{r,\theta,\phi}\paren{\frac{\partial N}{\partial t} \frac{u}{J}},
\end{equation}
minimized over the simulation domain. $N$ is the number of emitted
Monte Carlo packets, $u$ is gas internal energy density by volume, and
$J$ is the total frequency and angle integrated neutrino
emissivity. $Q_{\text{rad}}$ roughly encodes how well resolved the
radiation field is, with $Q_{\text{rad}}=10$ a marginal value. In our
simulation, we find $Q_{\text{rad}} \gtrsim 100$ for all time.

\section{Results}
\label{sec:results}

% \begin{figure}[tb!]
%   \centering
%   \includegraphics[width=\linewidth]{random-walk-conceptual}
%   \caption{A conceptual sketch of our definition of the average
%     advection time $\tau_a$ and the average advection velocity
%     $v_a$. At a given radius $r$, a Lagrangian fluid packet takes a
%     time $\Delta t$ to travel a height $\Delta z$. For $\Delta z= H$,
%     where $H$ is the scale height, we define $\tau_a := \Delta t$ and
%     $v_a = \Delta z/\Delta t$.}
%   \label{fig:random:walk:conceptual}
% \end{figure}

\subsection{Time Evolution and the Different Phases of Accretion}
\label{sec:time:evolution}

\begin{figure}[tb!]
  \centering
  \includegraphics[width=\linewidth]{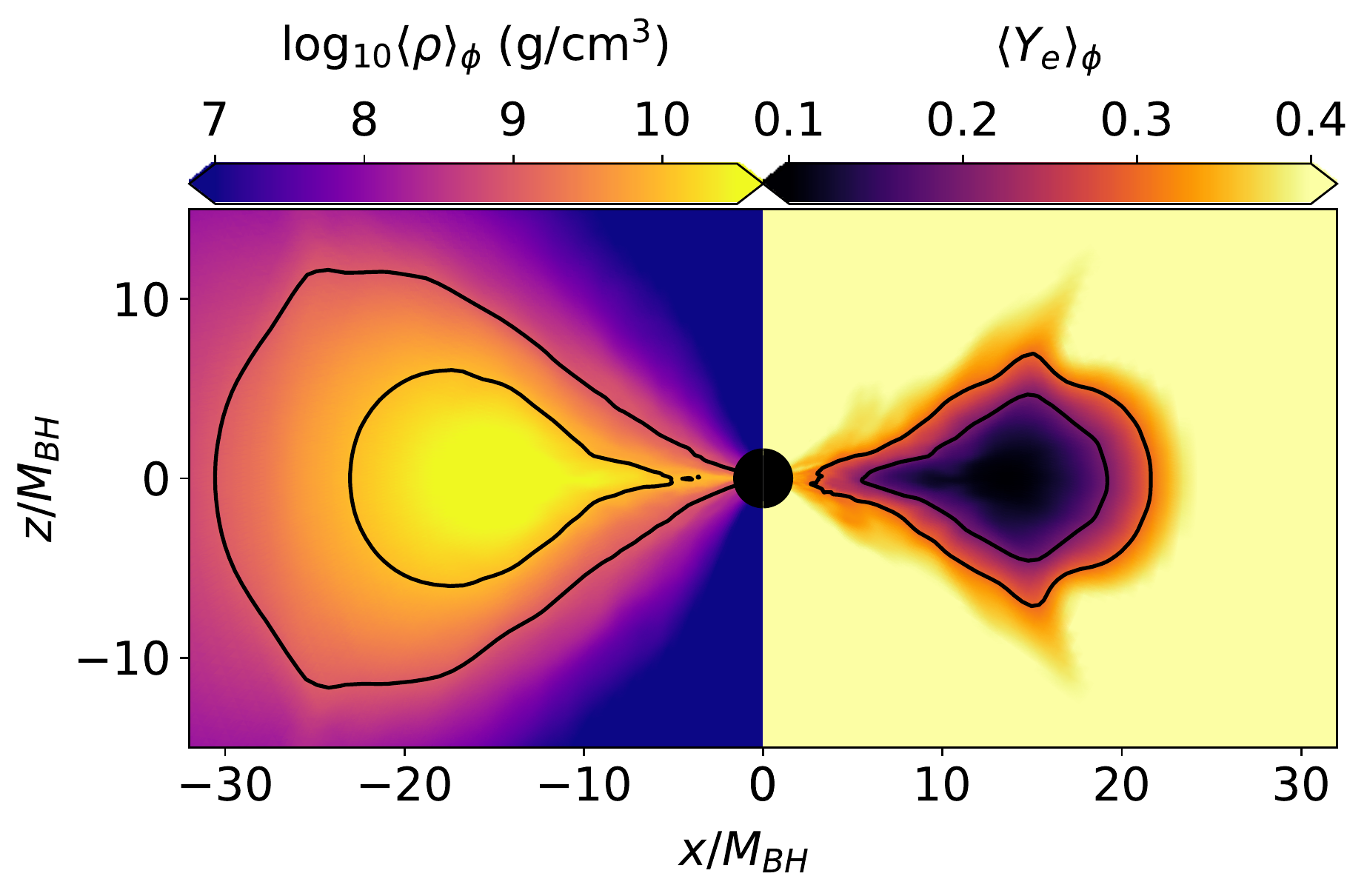}
  \caption{Same as Figure \ref{fig:rho:ye:t0} but at
    $t = 500 G M_{\rm BH}/c^3$ or $\approx 7$ ms.}
  \label{fig:rho:ye:t0500}
\end{figure}

\begin{figure}[tb!]
  \centering
  \includegraphics[width=\linewidth]{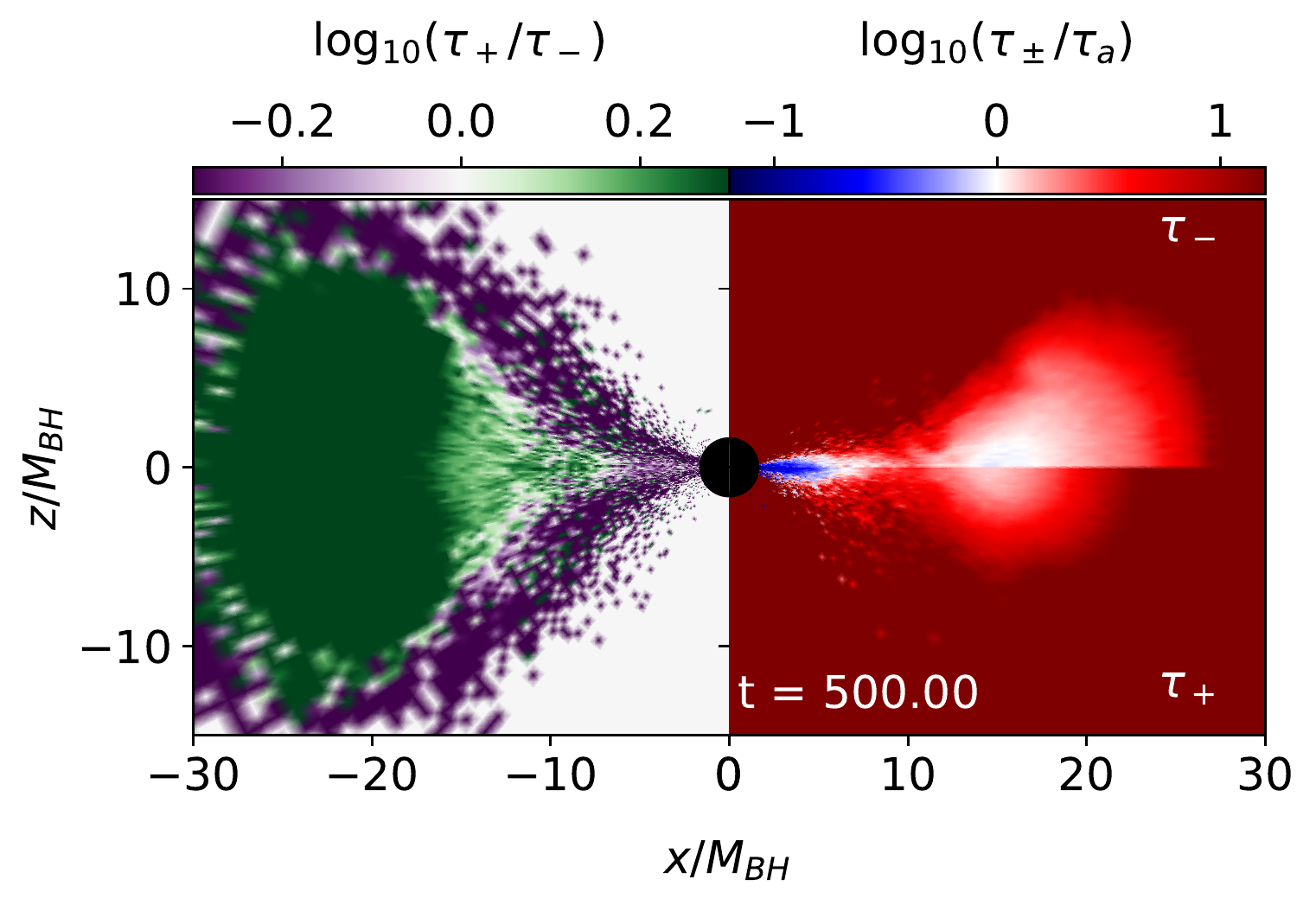}
  \caption{Same as Figure \ref{fig:time:scales:t0} but at
    $t = 500 G M_{\rm BH}/c^3$ or $\approx 7$ ms.}
  \label{fig:time:scales:t0500}
\end{figure}

The bottom pane of Figure \ref{fig:time:evolution} shows the time
evolution of the accretion rate of the disk. The accretion rate
matches the standard time profile for MRI-powered disks with compact
torus initial conditions \citep{Shiokawa2011,EHTComparison}. The
maximum accretion rate achieved for this realization is approximately
$10^{-1} M_\odot/s$, which eventually undergoes power-law decay. This
power law is well understood as an artifact of disks formed from a
finite reservoir of material. In Nature, the disk is fed by fallback
from the core-collapse event and thus the accretion rate will look
different. In the top pane, we plot the density-weighted mean
\begin{equation}
  \label{eq:ye:density:weighted:mean}
  \dwmean{Y_{\rm e}} = \frac{\int_\Omega \sqrt{-g} dx^3 \rho Y_{\rm e}}{\int_\Omega \sqrt{-g} dx^3 \rho},
\end{equation}
and standard deviation
\begin{equation}
  \label{eq:ye:density:weighted:std}
  \dwstd{Y_{\rm e}} = \sqrbrace{\frac{\int_\Omega \sqrt{-g} dx^3 \rho (Y_{\rm e} - \dwmean{Y_{\rm e}})^2}
    {\int_\Omega \sqrt{-g} dx^3 \rho}}^{1/2}
\end{equation}
of the electron fraction $Y_{\rm e}$ as functions of time, where
$$\int_\Omega \sqrt{-g}dx^3$$
represents an integral over the whole domain with appropriate measure.

The accretion rate, together with the evolution of the electron
fraction, suggest three phases, which will be described in more detail
below:
\begin{enumerate}[(a)]
\item As the initial torus disrupts and accretion flow is established,
  the disk rapidly deleptonizes. As a consequence, the mean electron
  fraction drops dramatically, while the standard deviation
  rises. This phase lasts for $t \lessapprox 500\ M_{\rm BH}$ or
  $t\lessapprox 7$ ms. We call this phase the \textit{initial
    transient} phase.
\item After accretion begins, the disk enters a short transition
  period before establishing a quasi-stationary flow. We call this
  phase the \textit{transition} phase, which roughly lasts for
  $500\ M_{\rm BH} \lessapprox t \lessapprox 1500\ M_{\rm BH}$ or
  $7\text{ ms}\lessapprox t \lessapprox 22\text{ ms}$.
\item In the quasi-stationary state, the accretion rate slowly drops
  as a power law.  As the accretion rate drops, the electron fraction
  slowly rises and the standard deviation slowly drops. We call this
  the \textit{steady-state} or \textit{quasi-stationary} phase, which
  lasts for $t \gtrapprox 1500\ M_{\rm BH} \approx 22\text{ ms}$.
\end{enumerate}
We emphasize that only phase \textbf{(c)} is representative of a
collapsar in Nature. Phases \textbf{(a)} and \textbf{(b)} are
unphysical transients that emerge from the initial
conditions. Nevertheless, a good understanding of these transients is
necessary for a complete picture of the structure in the disk in phase
\textbf{(c)}.

\begin{figure}[tb]
  \centering
  \includegraphics[width=\linewidth]{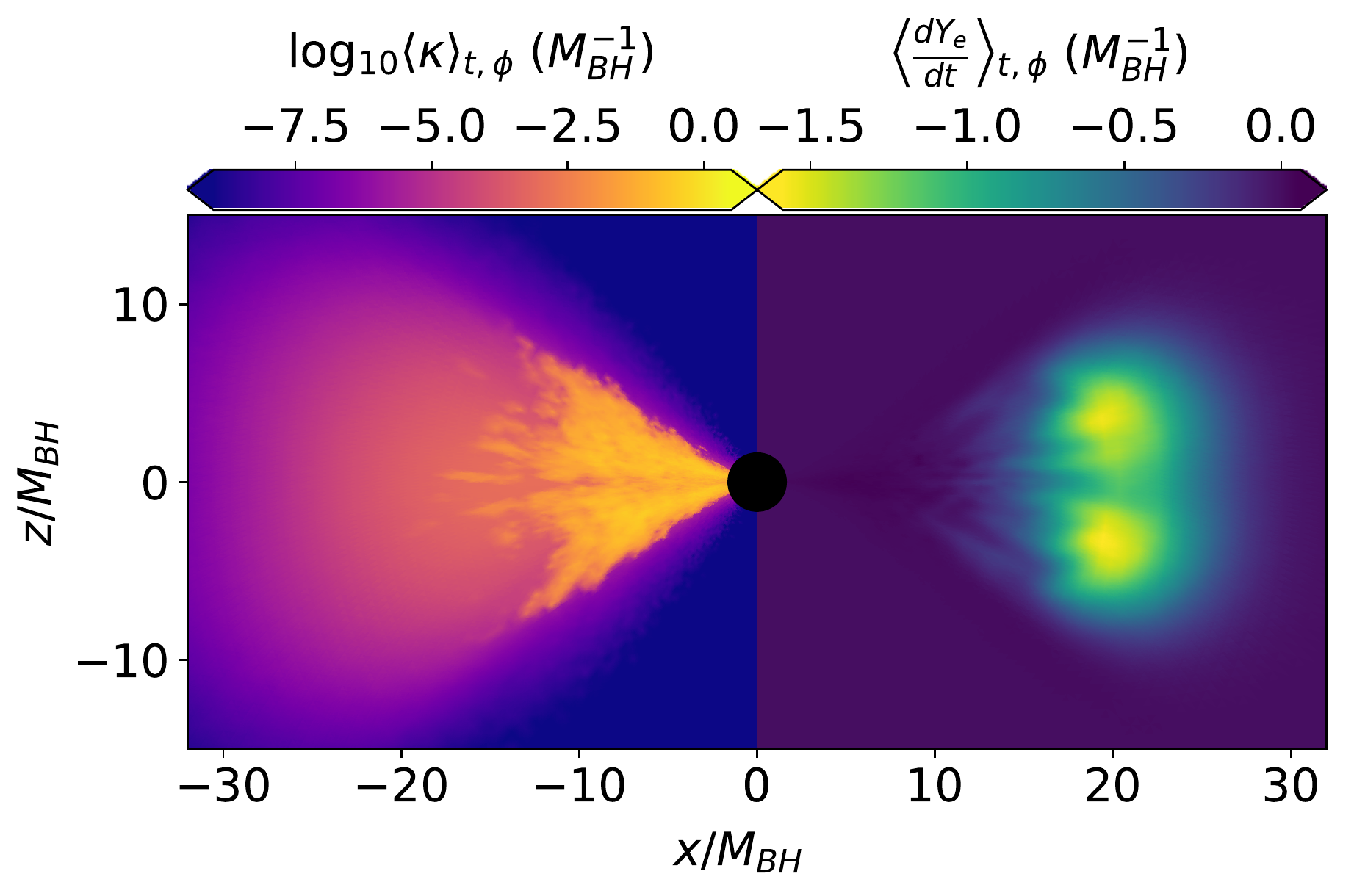}
  \caption{Neutrino opacity (left) and Lagrangian derivative of
    electron fraction in disk material (right). Opacity is integrated
    over frequency and flavor, weighted by the neutrino spectrum
    realized in the simulation. Both quantities are averaged over
    azimuthal angle $\phi$ and time in the early transient phase, from
    roughly 5 ms to 20 ms.}
  \label{fig:dtau:transient}
\end{figure}

The bottom pane of Figure \ref{fig:nu:luminosity} shows the luminosity
measured at infinity of electron neutrinos $\nu_e$, electron
antineutrinos $\bar{\nu}_e$, and all other neutrino species, grouped
together as ``heavy neutrinos'' or $\nu_x$. The top pane shows the raw
number of physical particles per second. The luminosity for heavy
neutrinos is universally orders of magnitude lower than for the
electron neutrinos and their antiparticles. The evolution of the
luminosity is consistent with phases \textbf{(a)}, \textbf{(b)}, and
\textbf{(c)}. At early times, the luminosity for electron neutrinos is
much higher than for electron anti-neutrinos, consistent with rapid
deleptonization. At late times, electron neutrinos and their
antiparticles are roughly in balance, with a slight excess in
anti-neutrinos, consistent with a slow releptonization process.

\subsection{Weak Equilibrium}
\label{sec:weak:equilibrium}

When the probability of $Y_{\rm e}$ increasing in packet of material is in
balance with the probability of $Y_{\rm e}$ decreasing, material is said to
be in \textit{weak equilibrium}. The classic example of weak equilibrium
is $\beta-$\textit{equilibrium} in a cold neutron star,
where the probability of $\beta$-decay
\begin{equation}
  \label{eq:def:beta:decay}
  n\to p + e + \bar{\nu}_e
\end{equation}
is in balance with inverse $\beta$-decay
\begin{equation}
  \label{eq:def:inverse:beta:decay}
  e + p \to n + \nu_e.
\end{equation}
See, e.g., \citet{shapiro2008black} for a detailed discussion. For
finite temperature systems out of equilibrium, such as the disk
discussed here, $\beta$-decay is too slow to be dynamically
relevant. Rather, the charged-current processes in table
\ref{tab:emis:abs} determine the equilibrium \citep{Freedman1974}.

For a given density $\rho$ and temperature $T$, we can calculate the
weak equilibrium electron fraction $Y_{\rm e}^{\rm eq}$ such that
these processes are in balance.\footnote{This calculation can be done
  in post-processing using the tabulated neutrino emissivities and
  opacities used by the code. For the purposes of
  $Y_{\rm e}^{\rm eq}$, and only $Y_{\rm e}^{\rm eq}$, we assume a
  radiation field in equilibrium with the gas.} The dotted black line
in the top pane of Figure \ref{fig:time:evolution} shows the
density-weighted mean value of $Y_{\rm e}^{\rm eq}$. Weak equilibrium
is an extremely strong assumption, not realized in the
simulation. However it provides a useful limiting case.

Throughout the accretion history, the disk is out of
equilibrium and neutrino processes drive it towards equilibrium. This
equilibrium is a moving target, as $Y_{\rm e}^{\rm eq}$ at a given
point in the disk changes as the conditions in the disk change with
the accretion rate, and is not achieved during the lifetime of the
simulation. $Y_{\rm e}$ begins far above the equilibrium value, which
drives the rapid deleptonization of phase \textbf{(a)}. As the
accretion rate falls, the equilibrium electron fraction
$Y_{\rm e}^{\rm eq}$ rises, which in turn drives the late-time
re-leptonization of the disk in phase \textbf{(c)}.

\begin{figure}[t!]
  \centering
  \includegraphics[width=\linewidth]{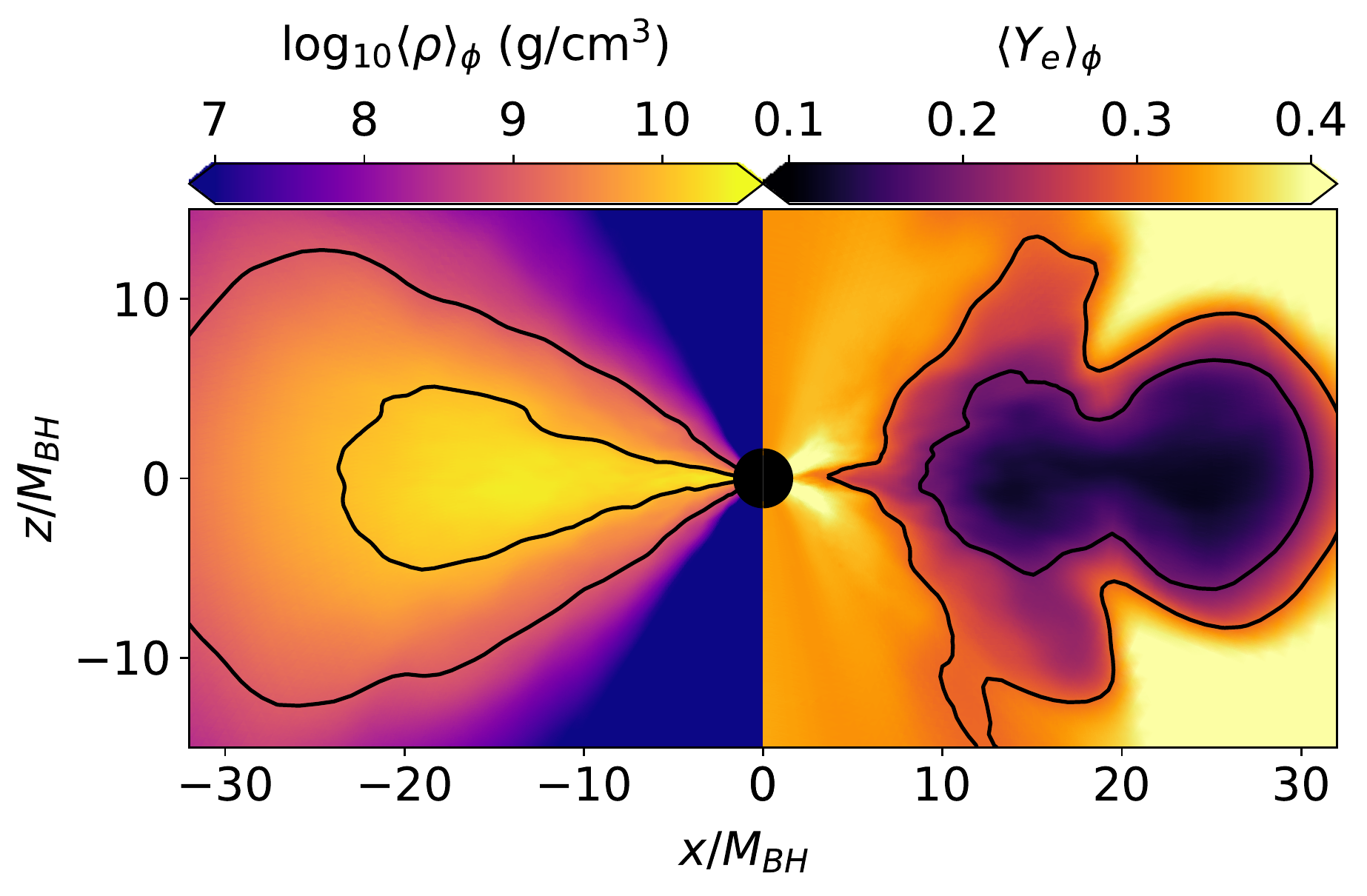}
  \caption{Same as Figure \ref{fig:rho:ye:t0} but at
    $t=1500\ G M_{\rm BH}/c^3$, or $\approx 22$ ms.}
  \label{fig:rho:ye:snap:t1500}
\end{figure}

\subsection{Relevant Time Scales}
\label{sec:relevant:time:scales}

Before we describe phases \textbf{(a)}, \textbf{(b)}, and \textbf{(c)}
in more detail, we introduce several concepts we will make use of
later. In particular, we are most interested in how electron fraction
$Y_{\rm e}$ is set within the disk. There are essentially three relevant
time scales. The first two are
\begin{equation}
  \label{eq:def:tau:+:-}
  \tau_\pm = \frac{\rho Y_{\rm e}}{G_{Y_{\rm e}}^\pm}
\end{equation}
where $G_{Y_{\rm e}}^\pm$ is the right-hand-side contribution to equation
\eqref{eq:lepton:cons} that increases (for $+$) or decreases (for $-$)
$Y_{\rm e}$ due to neutrino emission and absorption. Expressed more
succinctly,
\begin{equation}
  \label{eq:def:G:pm}
  G_{Y_{\rm e}} = G_{Y_{\rm e}}^+ - G_{Y_{\rm e}}^-.
\end{equation}
We calculate this in-line within \nubhlight~ by tracking the rate of
Monte Carlo particles of each species emitted or absorbed. For more
details see section 3 and equation 36 in \citet{nubhlight}.

\begin{figure}[tb!]
  \centering
  \includegraphics[width=\linewidth]{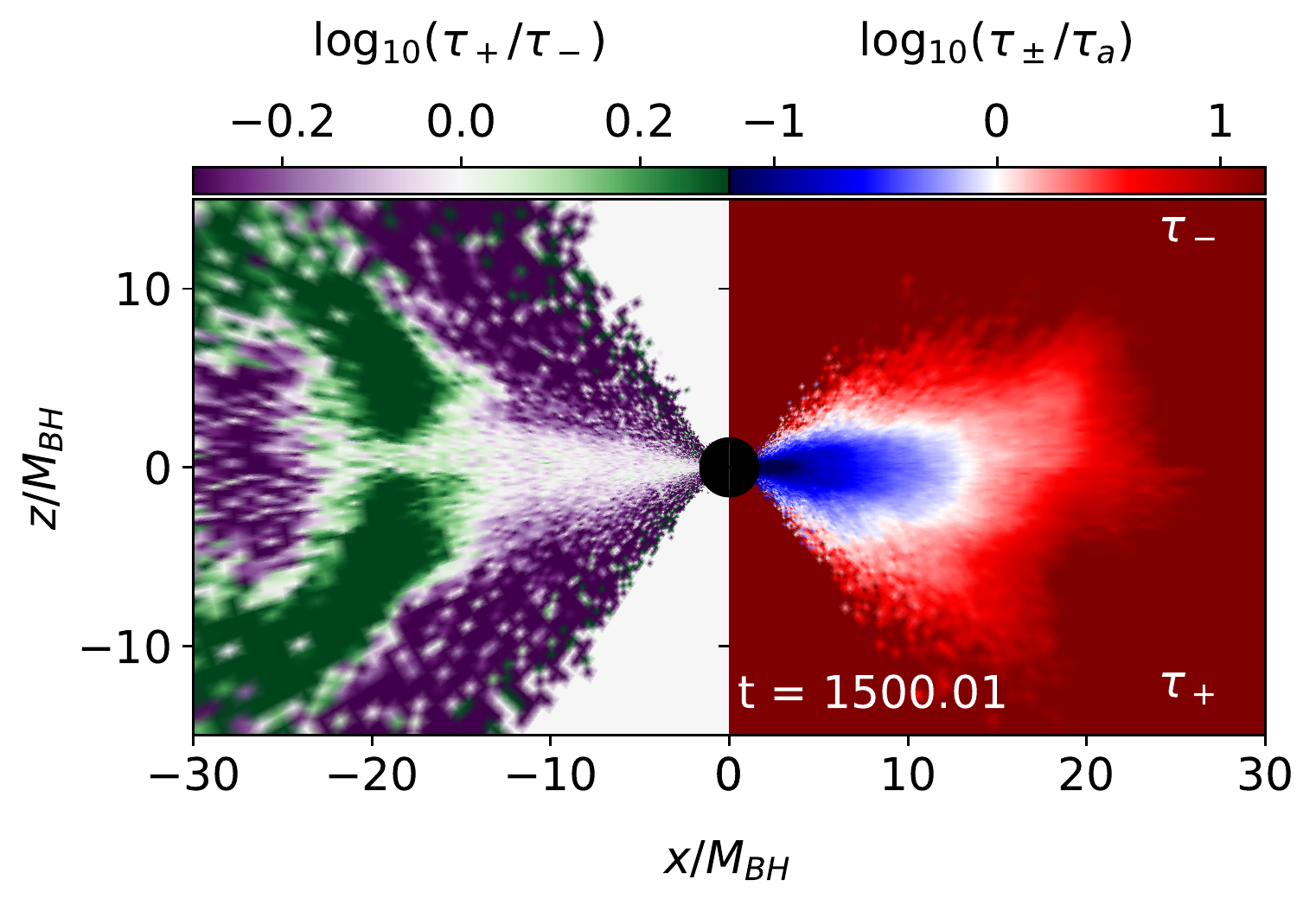}
  \caption{Same as Figure \ref{fig:time:scales:t0} but at
    $t = 1500 G M_{\rm BH}/c^3$ or $\approx 22$ ms.}
  \label{fig:time:scales:t1500}
\end{figure}

The third time scale is set by the rate at which electron fraction is
changed by advection and mixing. In other words, how long it takes for
a fluid packet with a given $Y_{\rm e}$ to be advected with the flow to
another location within the disk. As will be described in sections
\ref{sec:1D:model} and \ref{sec:outflow}, the electron fraction in the
disk is essentially a function of latitude $|90^\circ
-\theta|$. Therefore, we focus on advection in the $\theta$
direction. Extracting this time scale is complicated by the fact that
flow of material through the disk is not laminar. Indeed, MRI-driven
turbulence is the dominant momentum transport mechanism in the disk
\citep{BalbusHawley91,ShakuraSunyaevAlpha}.

Figure \ref{fig:selected:traces:latitude} shows how this fact
influences the trajectories of individual Lagrangian fluid packets. We
plot a random selection of tracer particles, which are advected with
the flow, that vary in latitude within the disk and eventually become
gravitationally unbound. The left column shows height $z$ as a
function of time, while the right column shows it as a function of
cylindrical radius
\begin{equation}
  \label{eq:def:cyl:rad}
  r_{\rm cyl}=\sqrt{x^2 + y^2}.
\end{equation}
Notice that tracer paths are not linear. Rather tracers, and thus
Lagrangian fluid packets, ``random walk'' through the space. Note that
the paths in Figure \ref{fig:selected:traces:latitude} project out the
azimuthal flow. Thus the random walk includes the meandering visible
in the figure and a turbulent orbit in the azimuthal direction.

To capture this idea, we introduce the spherically averaged, density
and $Y_{\rm e}$ weighted, \textit{advection velocity}
\begin{equation}
  \label{eq:def:adv:velocity}
  \left\langle v_a\right\rangle_{\rho,Y_{\rm e},\theta,\phi}(t,r) = \frac{\int_{S^2} \sqrt{-g} d^2x \rho Y_{\rm e} u^2}{\int_{S^2} \sqrt{-g} d^2x \rho Y_{\rm e}},
\end{equation}
where $u^2$ is the component of the four-velocity in the $\theta$
direction and
$$\int_{S^2}\sqrt{-g}d^2x$$
is an integral over a thin spherical shell with appropriate
measure. The integrand in the numerator of equation
\eqref{eq:def:adv:velocity} is the flux in the equation for the
conservation of lepton number \eqref{eq:lepton:cons}, while the
integrand in the denominator is general relativistically conserved
lepton number in the same equation. Equation
\eqref{eq:def:adv:velocity} thus tells us the rate that
\textit{leptons}, as opposed \textit{baryons} change their angle
$\theta$.

To transform this into a time scale, we need a characteristic length
scale. We follow \citet{Shiokawa2011} and calculate the \textit{scale
  angle} of the disk as a function of radius and time:
\begin{equation}
  \label{eq:def:thd}
  \theta_d(t,r) = \sqrt{\frac{ \int_{S^2} \sqrt{-g} d^2x \rho \theta^2}{\int_{S^2} \sqrt{-g} d^2x \rho}}
\end{equation}
and the \textit{scale height}
\begin{equation}
  \label{eq:def:H}
  H(t,r) = r\tan(\theta_d(t,r)).
\end{equation}
Then we define the \textit{average advection time}
\begin{equation}
  \label{eq:def:tau:a}
  \tau_a(r) =\frac{1}{t_f - t_i}\int_{t_i}^{t_f}dt \frac{\theta_d}{\left\langle v_a\right\rangle_{\rho,Y_{\rm e},\theta,\phi}},
\end{equation}
where we have introduced a time-average over the simulation time to
reduce noise. Note the units of equations \eqref{eq:def:adv:velocity}
and \eqref{eq:def:thd}. The advection velocity as we have defined it
has units of angle per time. Thus equation \eqref{eq:def:tau:a} has
units of time. An equivalent formulation could be constructed in terms
of scale heights, but would introduce an additional coordinate
transformation, as \nubhlight~uses approximately spherical
coordinates.

We emphasize that the advection time $\tau_a$ incorporates the
\textit{total} time it takes for a fluid packet to change
lattitude. This includes both the ``random walk'' in $z$ and the
azimuthal motion of the disk orbit. The advection velocity
\eqref{eq:def:adv:velocity} incorporates only motion in the $\theta$
direction, not the total speed of a fluid packet. Thus the increased
time a packet takes to change lattitude due to the fact that it is
orbiting the black hole and motion is not entirely vertical is
incorporated.

We will refer to $\tau_\pm$ and $\tau_a$ repeatedly
throughout the text. We note that these quantities can also be
computed more directly via an analysis of tracer particles, at the
price that performing a time-dependent analysis becomes more
difficult. We present a comparison between these two approaches below
in section \ref{sec:1D:model}.

% Once a steady-state flow is established and for small radii
% (say $r/M_{\rm BH} \lessapprox 30$), $\theta_d$ is roughly constant and
% $H\sim r$. See disk literature such as \citet{Shiokawa2011} or
% \citet{EHTComparison} for more details.

\subsection{The Initial Transient}
\label{sec:initial:transient}

We now discuss phase \textbf{(a)}. Figure \ref{fig:rho:ye:t0} shows a
moment very close to the initial time, at $t = 25 G M_{\rm BH}/c^3$ or
$\approx 0.3$ ms. Contours are for $\rho=10^9$ and $10^{10}$
g$/$cm$^3$. Although the initial torus is in \textit{hydrostatic}
equilibrium (which will be disrupted as the MRI develops), it is
\textit{not} in \textit{weak} equilibrium. Figure
\ref{fig:time:scales:t0} demonstrates this. We plot of $\tau_+/\tau_-$
(left), the ratio of $\tau_-/\tau_a$ (top right), and the ratio of
$\tau_+/\tau_a$ (bottom right). (Color bars are artificially saturated
to enhance ease of visualization.) The fact that $\tau_-$ is much
shorter than $\tau_+$ and $\tau_a$ indicates rapid electron capture
and subsequent rapid decrease in $Y_{\rm e}$.

\subsection{The Transition Phase}
\label{sec:transition:phase}

We now discuss phase \textbf{(b)}. Figure \ref{fig:rho:ye:t0500} shows
the density and electron fraction at roughly the beginning of this
phase, $t = 500 G M_{\rm BH}/c^3$ or $\approx 7$ ms. Figure
\ref{fig:time:scales:t0500} shows the time scales $\tau_+$, $\tau_-$,
and $\tau_a$. As evidenced by the ratio of $\tau_+/\tau_-$, the disk
is still deleptonizing, but the core of the disk has achieved very low
electron fractions---as low as $Y_{\rm e}\sim 0.15$, similar to that
in the neutron star merger disk case \citep{MillerGW170817} and
consistent with \citet{SiegelCollapsar}.

At this point the relativistic, highly-magnetized jet of material
characteristic of MRI-driven accretion is beginning to develop. The
jet is visible as the absence of material in the polar regions in the
left pane of Figure \ref{fig:rho:ye:t0500}. In our simulation, Baryon
loading from the artificial atmosphere required by the Eulerian nature
of the simulation \citep[c.f.][]{nubhlight} prevents the jet from
reaching the very large Lorentz factors realized in Nature. However,
the region is highly magnetized and the jet is powerful enough to
evacuate the region.

\begin{figure}[t!]
  \centering
  \includegraphics[width=\linewidth]{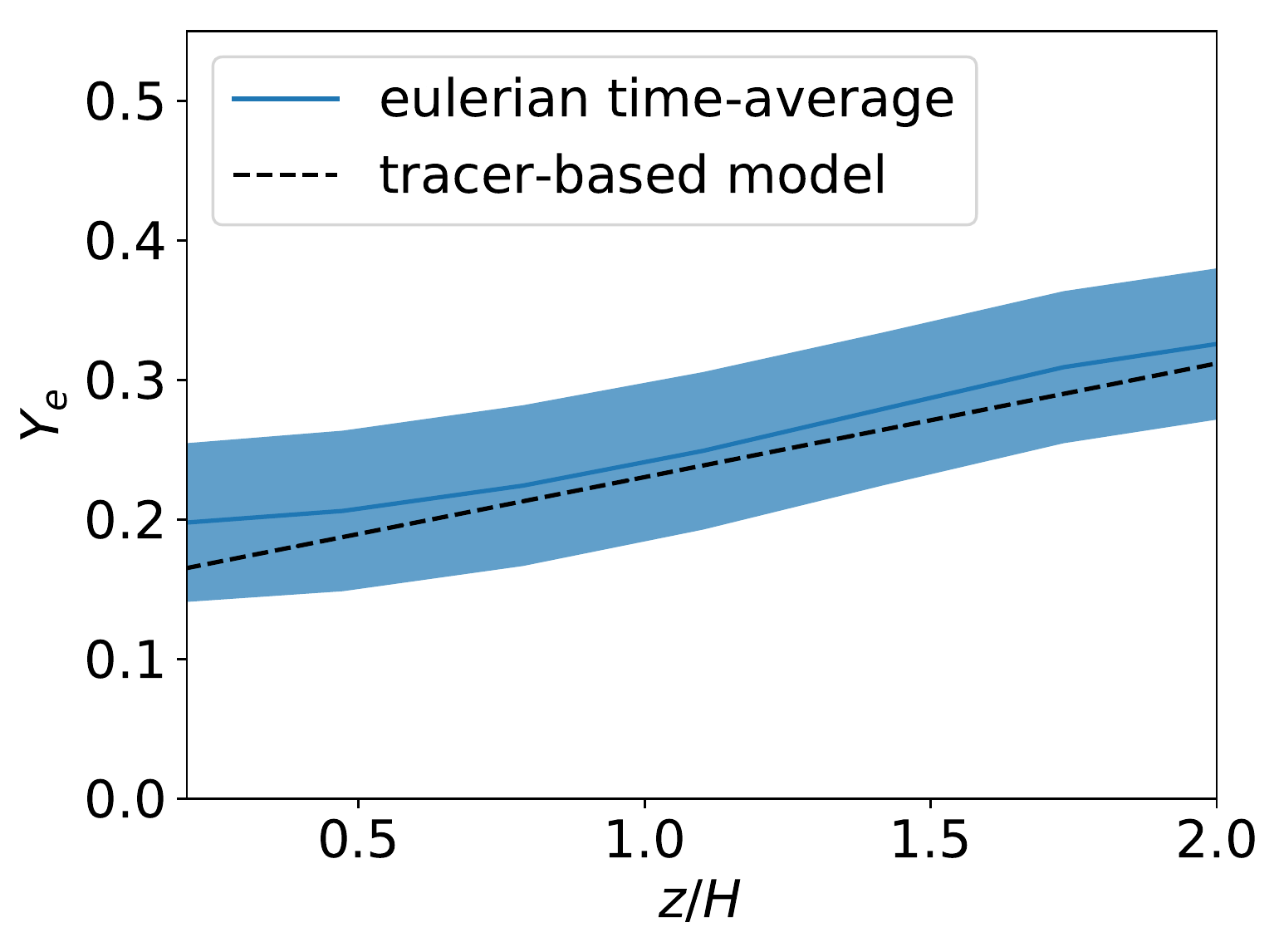}
  \caption{Vertical structure of the electron fraction, $Y_{\rm e}$ as
    a function of $z/H$, in the disk computed two ways. The solid line
    and envelope show the mean and standard deviation of the electron
    fraction averaged over the disk and over time, as computed in
    equations \eqref{eq:def:Ye:zoh:mean} and
    \eqref{eq:def:Ye:zoh:std}. The dashed line is computed using our
    simple 1D model \eqref{eq:1D:model} calibrated to tracer data.}
  \label{fig:1D:fit:test}
\end{figure}

\subsection{The Importance of Absorption at Early Times}
\label{sec:absorption:early:time}

In phases \textbf{(a)} and \textbf{(b)}, both neutrino absorption and
emission matter for setting the electron fraction. One way of
characterizing how much absorption matters is the neutrino optical
depth $\tau$. $\tau \ll 1$ implies a free-streaming limit, while
$\tau \gg 1$ implies a diffusion limit
\citep{castor2004radiation}.\footnote{Note that the relative volume
  density of leptons in the gas also matters.}

Figure \ref{fig:dtau:transient} shows both the rapid deleptonization
of the disk and the mitigating effect due to absorption. We compute an
effective opacity $\braket{\kappa}$ by measuring the number of
neutrinos of each flavor and frequency absorbed by the gas
\textit{in-situ} in \nubhlight. This amounts to integrating the
frequency and flavor dependent opacity over frequency, weighted by the
neutrino spectrum as realized in the simulation and summing over
flavor, again weighted by the abundances per flavor as realized in the
simulation. We plot this in the left pane of Figure
\ref{fig:dtau:transient}. The right pane shows the co-moving
derivative of electron fraction. As the disk disrupts, densities and
temperatures rise near the black hole, causing opacities, and thus
optical depths, to become significant, which prevents the electron
fraction from dropping as rapidly.

\subsection{Early-Time Quasi-Stationary Structure}
\label{sec:steady:state}

\begin{figure}[tb!]
  \centering
  \includegraphics[width=\linewidth]{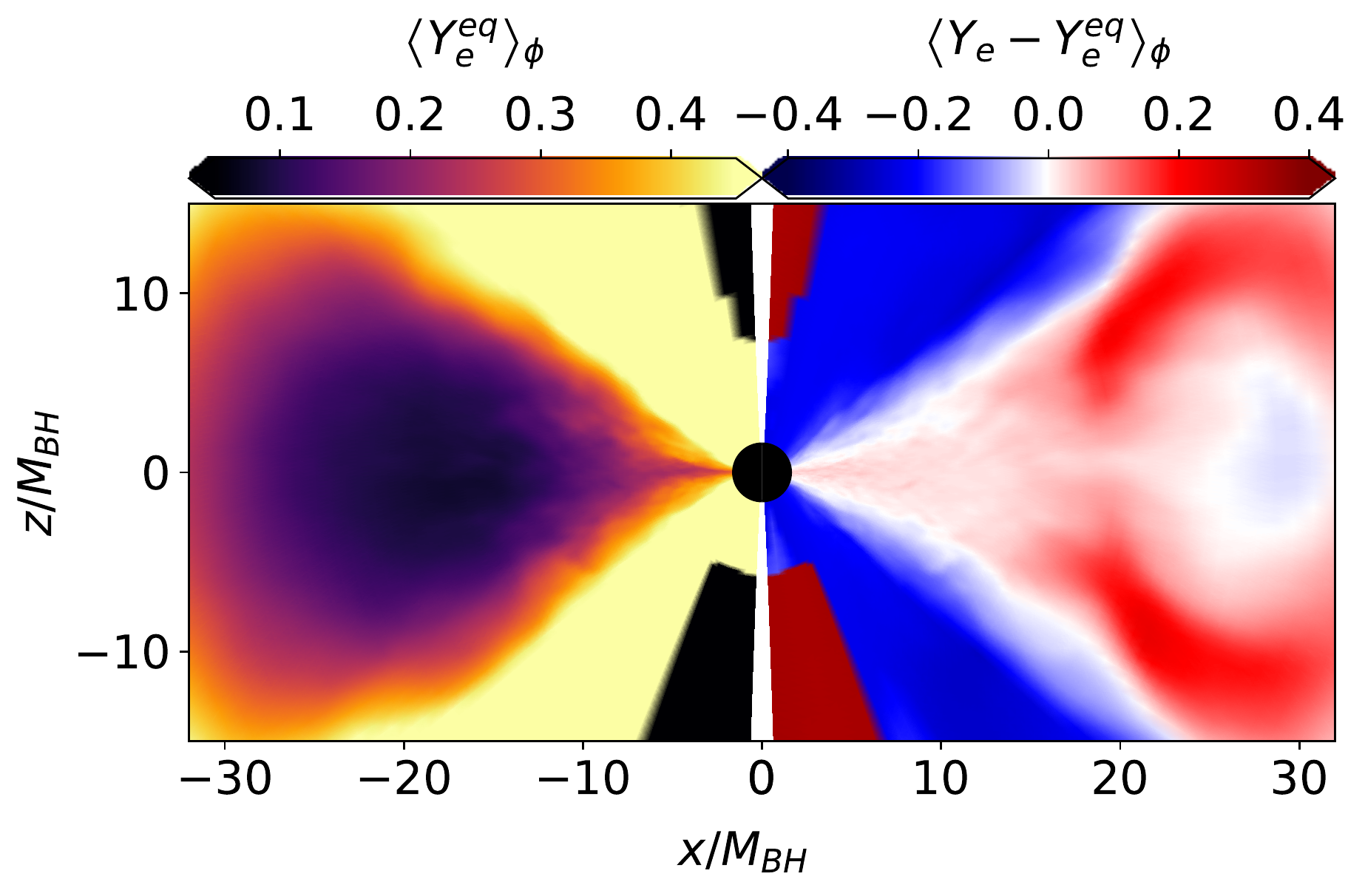}
  \caption{Equilibrium electron fraction $Y_{\rm e}^{\rm eq}$ (left) and the
    difference between the true electron fraction and the equilibrium
    value (right) at $t = 1500 G M_{\rm BH}/c^3$ or $\approx 22$ ms and
    averaged over azimuthal angle $\phi$. For
    $\sqrt{x^2+y^2} \lessapprox 20 M_{\rm BH}$, and near the equator, the electron
    fraction is close to the equilibrium value. However, at higher
    latitudes, $Y_{\rm e}^{\rm eq} > Y_{\rm e}$, driving a releptonization of material
    as it rises in latitude. }
  \label{fig:ye:equil:snap}
\end{figure}

\begin{figure}[t!]
  \centering
  \includegraphics[width=\linewidth]{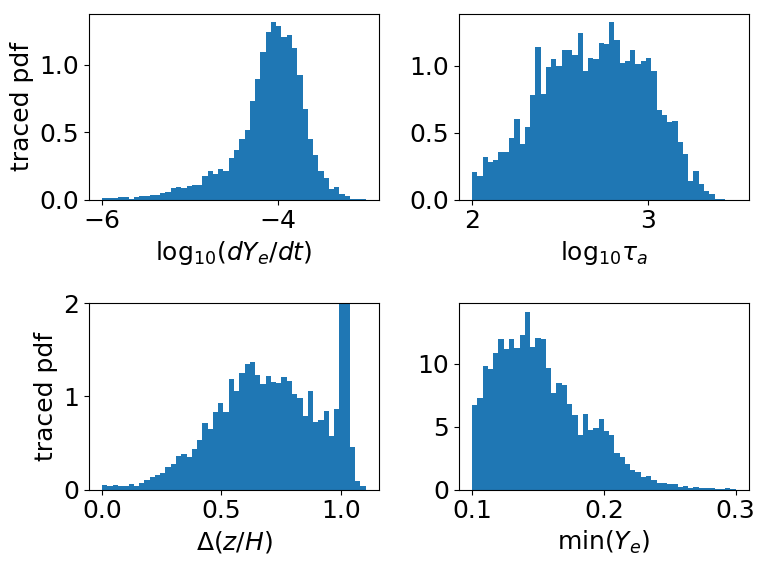}
  \caption{Probability distributions of the quantities used in
    equation \eqref{eq:1D:model} and binned from tracer particles. For
    this analysis, we use tracers that exist within the disk at least
    at time $t=1500 G M_{\rm BH}/c^3$ but that eventually escape.}
  \label{fig:fit:parameters}
\end{figure}

We now discuss phase \textbf{(c)}. As the accretion rate is slowly
decreasing in this stage, this steady-state is not truly steady, but
evolves adiabatically with time. As such we will highlight two times in this phase and describe how the system evolves adiabatically between them.

We begin with figure \ref{fig:rho:ye:snap:t1500}, which shows the
density and electron fraction of the disk at the beginning of phase
\textbf{(c)}, roughly at $t=1500\ G M_{\rm BH}/c^3$, or $\approx 22$
ms. The jet is now well-developed, as evidenced by a sharp drop in
density near the poles. Figure \ref{fig:time:scales:t1500} shows time
scales at the same time. At this point, the disk has finished
deleptonizing. Near the equator, $\tau_+$ and $\tau_-$ are roughly
equal and shorter than $\tau_a$, indicating the electron fraction is
stable in this region.

At slightly higher latitudes, for $r \lessapprox 15 G M_{\rm BH}/c^2$,
$\tau_+$, $\tau_-$, and $\tau_a$ are all roughly the same order of
magnitude. However, $\tau_+$ and $\tau_-$ are out of balance, with
$\tau_+$ shorter than $\tau_-$. We also observe that it is at this
point that a stratified structure in the electron fraction begins to
appear. For a given radius, $Y_{\rm e}$ is higher at higher latitudes.

We use the electron fraction averaged over radius and azimuthal angle
under a change of variables from $(r,\theta,\phi)$ to $(r,z/H,\phi)$
\begin{equation}
  \label{eq:def:Ye:zoh}
  \braket{Y_{\rm e}}_{r,\phi}(t,z/H) = \int_{r,\phi} \sqrt{-g} d^2x Y_{\rm e}(t,r,z/H(r),\phi)
\end{equation}
to define a measure of the vertical structure. We average this
quantity over time in this phase to get the mean
\begin{equation}
  \label{eq:def:Ye:zoh:mean}
  \braket{Y_{\rm e}}_{t,r,\phi}(z/H )= \frac{1}{\Delta t}\int dt \int_{r,\phi} \braket{Y_{\rm e}}_{r,\phi}(t,z/H)
\end{equation}
and standard deviation
%\begin{widetext}
%\begin{footnotesize}
\begin{equation}
  \label{eq:def:Ye:zoh:std}
  \text{std}\paren{Y_{\rm e}}_{t,r,\phi}(z/H) = \sqrt{\frac{\int dt \paren{\braket{Y_{\rm e}}_{r,\phi} - \braket{Y_{\rm e}}_{t,r,\phi}}^2}{\Delta t}}
\end{equation}
%\end{footnotesize}
%\end{widetext}
of the vertical structure. The solid line in figure
\ref{fig:1D:fit:test} shows the mean \eqref{eq:def:Ye:zoh:mean} and
the envelope shows the standard deviation \eqref{eq:def:Ye:zoh:std}.

\subsection{A Simple One-Dimensional Model}
\label{sec:1D:model}

We propose the following simple, one-dimensional model to explain this
structure. The left pane of Figure \ref{fig:ye:equil:snap} shows the
equilibrium electron fraction $Y_{\rm e}^{\rm eq}$ introduced in section
\ref{sec:weak:equilibrium} at the beginning of phase \textbf{(c)}, or
$t = 1500 G M_{\rm BH}/c^3 \approx 22$ ms. This is not the
\textit{physical} electron fraction $Y_{\rm e}$ attained in the
simulation. Rather, it is the value where weak processes are in
balance. The right pane shows the difference between the true $Y_{\rm e}$
and $Y_{\rm e}^{\rm eq}$. For $\sqrt{x^2+y^2} \lessapprox 20 M_{\rm BH}$ and near
the equator, $Y_{\rm e}$ is close to the equilibrium value. However, at
higher latitudes, $Y_{\rm e} < Y_{\rm e}^{\rm eq}$.

Material at these higher latitudes is fed by material closer to the
equator. As described in section \ref{sec:relevant:time:scales}, this
takes time as the material ``random walks'' from the lower latitudes
to higher latitudes.\footnote{Recall that the random walk includes not
  only vertical motion, but turbulent azimuthal motion as a fluid
  packet orbits the black hole.} As the material rises and the density
of the fluid decreases, the emission and absorption of neutrinos raise
the electron fraction. The electron fraction at a given height $z$
relative to the scale height $H$ is then given by the rate at which
$Y_{\rm e}$ is increasing in time due to weak processes times the
amount of time it took for the fluid to random walk to that height.

We use our tracer particles to implement this model as
\begin{widetext}
\begin{equation}
%\begin{eqnarray}
  \label{eq:1D:model}
  %Y_{\rm e}(z/H) &=& \braket{\text{min}(Y_{\rm e})}_{\text{trc}}%\\
  %&& + \braket{\frac{d Y_{\rm e}}{dt}}_{t,\text{trc}} \paren{H\braket{\frac{dz}{dt}}_{t,\text{trc}}^{-1}}\paren{\frac{z}{H} - \braket{\text{min}(z/H)}_{\text{trc}}}\nonumber
  Y_{\rm e}(z/H) = \braket{\text{min}(Y_{\rm e})}_{\text{trc}}%\\
  + \braket{\frac{d Y_{\rm e}}{dt}}_{t,\text{trc}} \paren{H\braket{\frac{dz}{dt}}_{t,\text{trc}}^{-1}}\paren{\frac{z}{H} - \braket{\text{min}(z/H)}_{\text{trc}}}
%\end{eqnarray}
\end{equation}
\end{widetext}
where we have assumed
\begin{displaymath}
  \braket{\frac{d(z/H)}{dt}}_{t,\text{trc}}^{-1} \approx H\braket{\frac{dz}{dt}}_{t,\text{trc}}^{-1},
\end{displaymath}
which is safe for slowly varying $H$ and where
\begin{equation}
  \label{eq:def:trc:average}
  \braket{Q}_{\text{trc}} = \frac{\sum_{i = 0}^{N_t} m_i Q_i}{\sum_{i=0}^{N_t} m_i}
\end{equation}
is a mass-weighted sum over $N_t$ tracer particles. We also define an
average over tracer history:
\begin{equation}
  \label{eq:def:trc:average}
  \braket{Q}_{t,\text{trc}} = \frac{\sum_{i = 0}^{N_t} m_i \int_{t_0}^{t_c} dt Q_i}{(t_c-t_0)\sum_{i=0}^{N_t} m_i},
\end{equation}
where $t_0$ is the time that the tracer achieves its minimum latitude
$z/H$ and $t_c > t_0$ is the time that the tracer then rises to a
latitude of one scale height $z=H$. Here we average over tracer
particles that are in the disk at $t = 1500 G M_{\rm BH}/c^3$ but
eventually become gravitationally unbound. (Here we define
``gravitationally unbound'' as reaching an extraction radius of 250
$GM_{\rm BH}/c^2$ with a positive Bernoulli parameter. More distant
choices of extraction radius do not change the results presented
here.)

\begin{figure}[tb!]
  \centering
  \includegraphics[width=\linewidth]{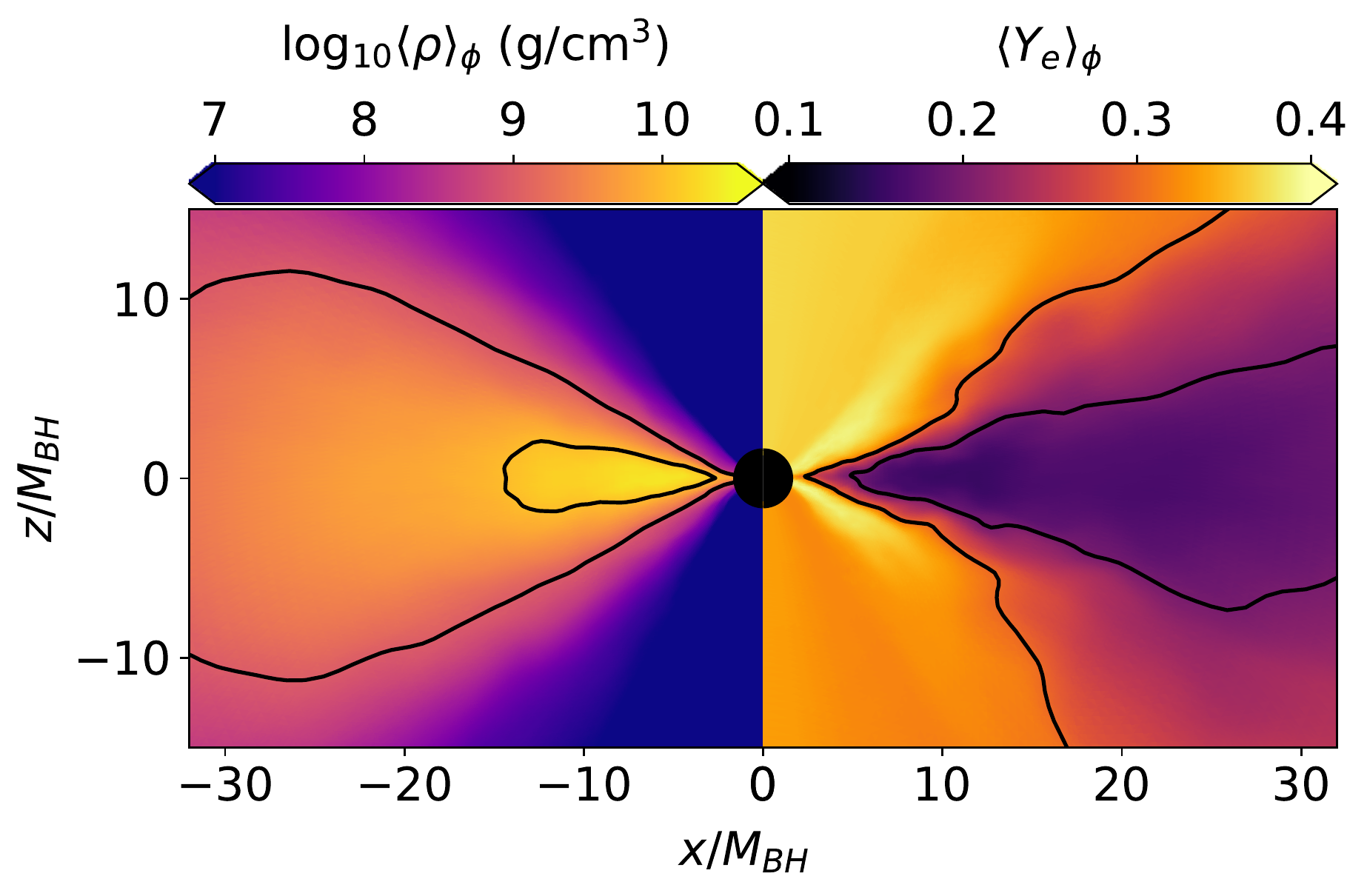}
  \caption{Same as Figure \ref{fig:rho:ye:t0} but at
    $t = 5000 G M_{\rm BH}/c^3$ or $\approx 73$ ms.}
  \label{fig:rho:ye:t5000}
\end{figure}

\begin{figure}[tb!]
  \centering
  \includegraphics[width=\linewidth]{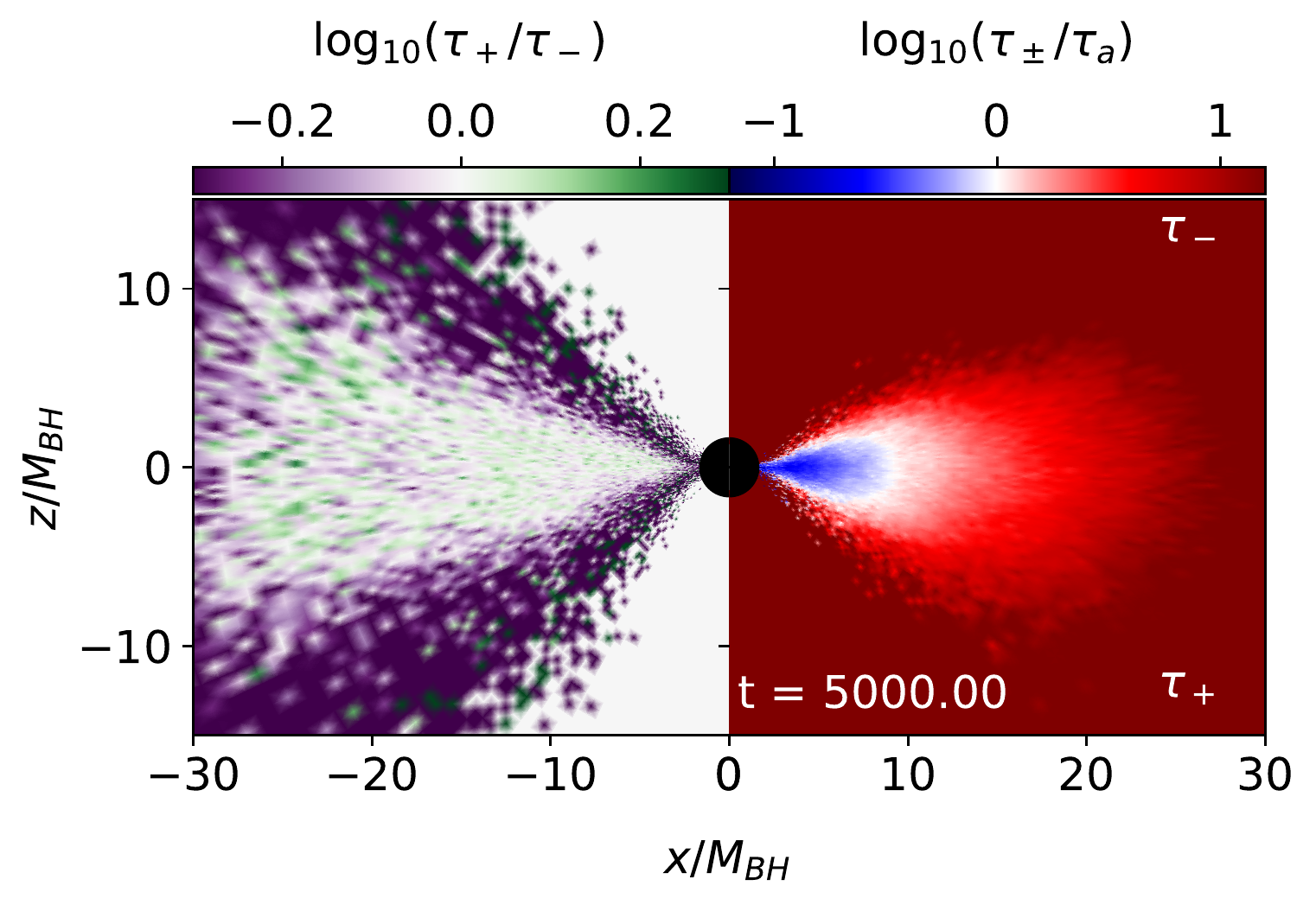}
  \caption{Same as Figure \ref{fig:time:scales:t0} but at
    $t = 5000 G M_{\rm BH}/c^3$ or $\approx 73$ ms.}
  \label{fig:time:scales:t5000}
\end{figure}

The probability distributions for the parameters for equation
\eqref{eq:1D:model} are binned from tracer particles in Figure
\ref{fig:fit:parameters}. We use the mean values. Figure
\ref{fig:1D:fit:test} compares our model \eqref{eq:1D:model} extracted
from tracers to the vertical structure of $Y_{\rm e}$ in the disk extracted
from an Eulerian picture of the flow. The dashed line is the 1D model
fit to the tracer data and the solid line and envelope are extracted
from the Eulerian picture via equations \eqref{eq:def:Ye:zoh:mean} and
\eqref{eq:def:Ye:zoh:std}. The true flow structure is not entirely
linear---it saturates at extreme latitudes. Nevertheless, the simple
linear model does remarkably well---agreement is well within the
standard deviation.

This stratification of the electron fraction has implications in the
outflow. Wind material that passes through these higher latitude
regions as it leaves the disk will have its electron fraction set at
least in part according to equation \eqref{eq:1D:model}. Indeed, it
will likely releptonize further as it achieves a greater distance from
the disk.

\subsection{Late Times}
\label{sec:late:times}

As the disk continues to accrete, the density and accretion rate
drop. Figure \ref{fig:rho:ye:t5000} shows the disk well into phase
\textbf{(c)}, at $t = 5000 G M_{\rm BH}/c^3$ or $\approx 73$ ms. Figure
\ref{fig:time:scales:t5000} shows the time scales $\tau_+$, $\tau_-$,
and $\tau_a$. As the density drops, the weak processes in the disk
both slowly drive the disk towards larger $Y_{\rm e}$ and slowly shut off as
the time scales $\tau_+$ and $\tau_-$ gradually become large beyond
dynamical relevance.

As the time scales $\tau_\pm$ rise, the turbulence in the disk is
better able to mix the disk, and the stratified structure described in
the earlier sections slowly homogenizes away. We are now in a position
to understand the evolution of the mean and standard deviation of
$Y_{\rm e}$ shown in figure \ref{fig:time:evolution}. The standard deviation
is a measure of stratification initially powered by weak processes and
slowly erased by turbulent mixing. If we ran the simulation for
longer, the accretion rate would continue to fall, the mean electron
fraction would continue to rise, and the electron fraction in the disk
would continue to homogenize until at very low accretion rates the
disk would become composed of symmetric matter.

% Figure \ref{fig:rho:ye:snap} shows the density $\rho$ and electron
% fraction $Y_{\rm e}$ in the disk after the transient cutoff of
% $5\times 10^3 G M_{\rm BH}/c^3$, which is about $74$ ms. The densities
% closest to the black hole are as large as $10^{10}$ g$/$cm$^3$. In the
% mid-plane, the disk attains low electron fraction, with
% $Y_{\rm e} \sim 0.15$, similar to that in the neutron star merger disk case
% \citep{MillerGW170817}. However, at higher latitudes, the electron
% fraction in the disk is higher.

\subsection{Outflow}
\label{sec:outflow}

% \begin{figure}[t!]
%   \centering
%   \includegraphics[width=\linewidth]{rho_Ye_snap}
%   \caption{Density $\rho$ (left) and electron fraction $Y_{\rm e}$ (right)
%     at $t=5\times 10^3 G M_{\rm BH}/c^3$, or $\approx 74$ ms. Contours are
%     for $\rho = 10^9, 10^{10}$ g$/$cm$^3$ and $Y_{\rm e} = 0.2, 0.3$
%     respectively. Both quantities are averaged over azimuthal angle
%     $\phi$.}
%   \label{fig:rho:ye:snap}
% \end{figure}

\begin{figure}[tb!]
  \centering
  \includegraphics[width=\linewidth,clip,trim={0 75 0 75}]{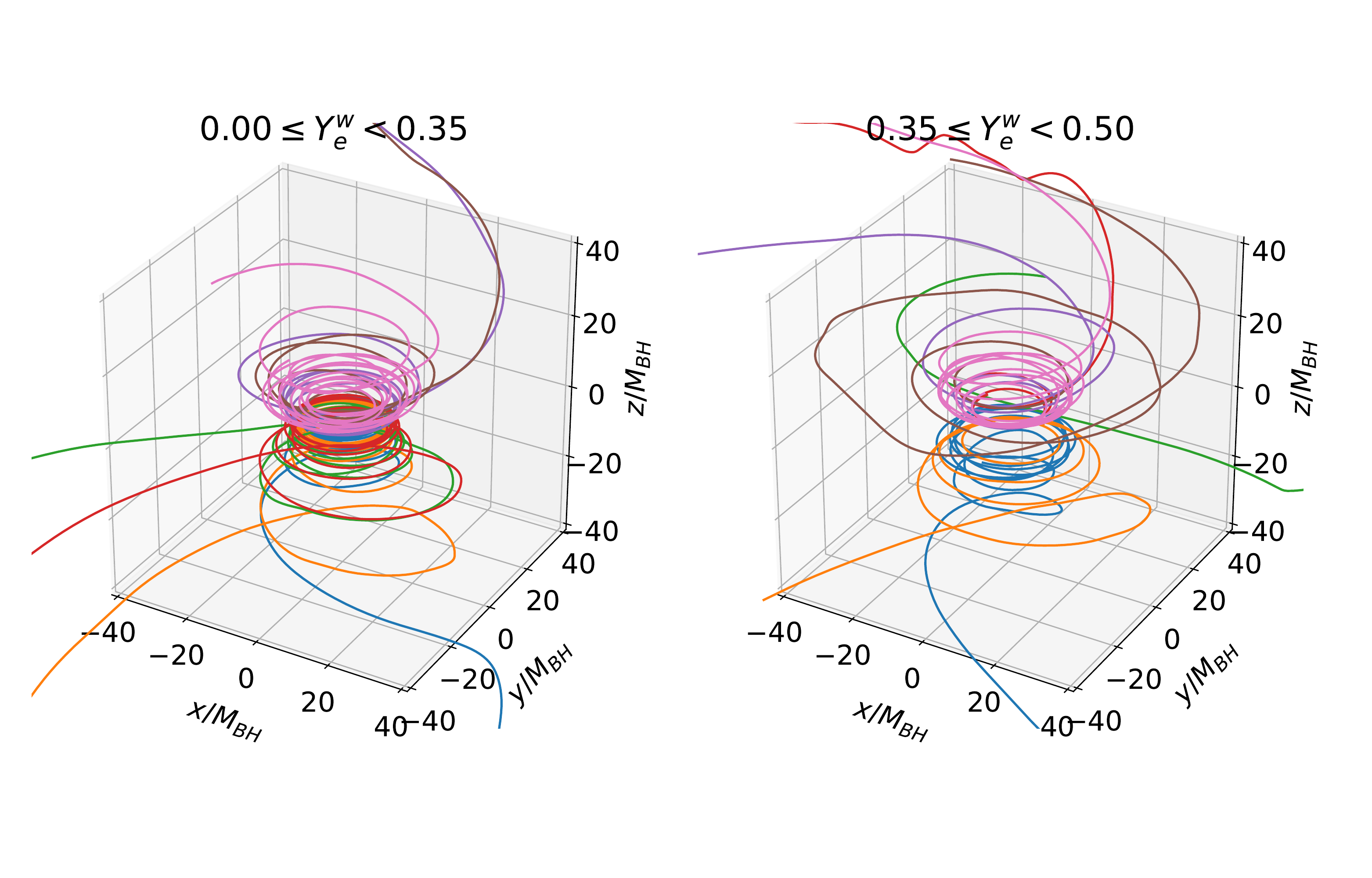}
  \caption{Paths of a selection of tracer particles in 3d. We split
    the tracers into those with $Y_{\rm e} < 0.35$ (left) and those with
    $Y_{\rm e} \geq 0.35$ (right). Broadly, tracers with lower electron
    fraction spend more time near the polar axis, while tracers with
    higher electron fraction spend less time. Colors highlight
    different traces to guide the eye.}
  \label{fig:tracerlines:3d}
\end{figure}

\begin{figure}[tbp]
  \centering
  \includegraphics[width=\linewidth]{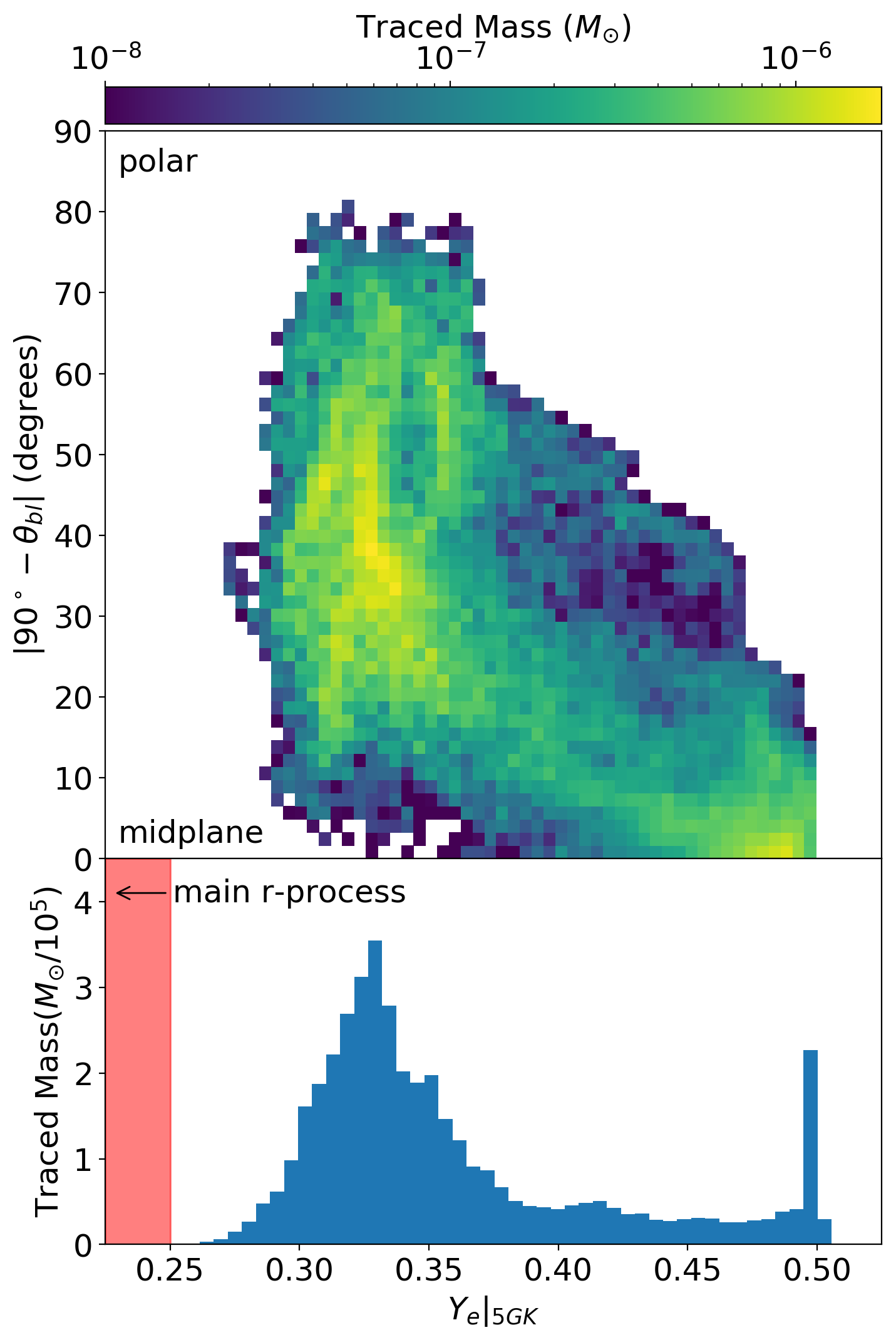}
  \caption{Electron fraction $Y_{\rm e}$ in outflow (top) vs angle and
    (bottom) binned by mass. The electron fraction is universally
    large, higher than $Y_{\rm e} > 0.25$. $Y_{\rm e}$ is lower for more polar
    outflow. The spike in $Y_{\rm e}\approx 0.5$ is from viscous spreading
    at the back of the disk, which never drops from its initial $Y_{\rm e}$
    to low electron fraction.}
  \label{fig:ye:v:theta}
\end{figure}

We now move our attention to disk outflow and implications for
nucleosynthesis. Figure \ref{fig:tracerlines:3d} shows the paths of a
selection of gravitationally unbound Lagrangian tracer particles. We
split the tracers into those with $Y_{\rm e} < 0.35$ and those with
$Y_{\rm e} \geq 0.35$. Qualitatively, we find that tracers with lower
electron fraction tend to spend more time close to the polar
axis. About 1 in 100 tracers have near-vertical trajectories, implying
they may be entrained in the jet or that they are interacting with the
funnel wall. The prospect of nucleosynthetic material entrained in the
jet has been explored in a number of works and is potentially
consistent with our results. See
\citet{Fujimoto2007,Ono2012,Nakamura2015,Soker2017,Kayakawa2018} for
some examples.

The electron fraction in the outflow is bounded from below by
$Y_{\rm e} \gtrsim 0.25$. The polar outflow has lower electron fraction than
the mid-plane outflow, as shown in figure \ref{fig:ye:v:theta}. This is
in contrast to the neutron star merger case, where the polar outflow
had \textit{higher} electron fraction than the mid-plane
\citep{MillerGW170817}.

% This may be related to the different
% initial conditions---in \citet{MillerGW170817}, the initial torus had
% $Y_{\rm e}=0.1$.

As the disk accretes, magnetically-driven turbulence transports mass
in the mid-plane radially inward and angular momentum radially
outward. Some material must carry this angular momentum to
infinity. The outflow driven by momentum conservation and turbulent
viscosity is sometimes referred to as the viscous spreading of the
disk. In the neutron star merger case, this viscous spreading is
physically meaningful; the disk is not fed, but rather develops from
material close to the black hole left over from the merger event. In
contrast, in the jet-driven supernova case, the disk is fed by
fallback material from the stellar envelope. For completeness, we
record this material and count it in our analysis. However, it is not
clear that mid-planar outflow will escape the star or that it is
physically meaningful.

Although the electron fraction is above $Y_{\rm e} \sim 0.25$, which is the
approximate threshold for robust r-process nucleosynthesis, entropy
can play a role in the nucleosynthetic yields as well. In particular,
high entropy material may undergo robust r-process even in a less
neutron-rich environment. For example, material with entropy of
$s=100 k_b/$baryon and $Y_{\rm e}=0.35$ may undergo a robust
r-process. High-velocity, shocked material entrained in the jet might
become high entropy. Magnetic reconnection in the jet may also drive
up entropy. Therefore, we investigate the velocity and entropy of the
outflow.

\begin{figure}[tbp]
  \centering
  \includegraphics[width=\linewidth]{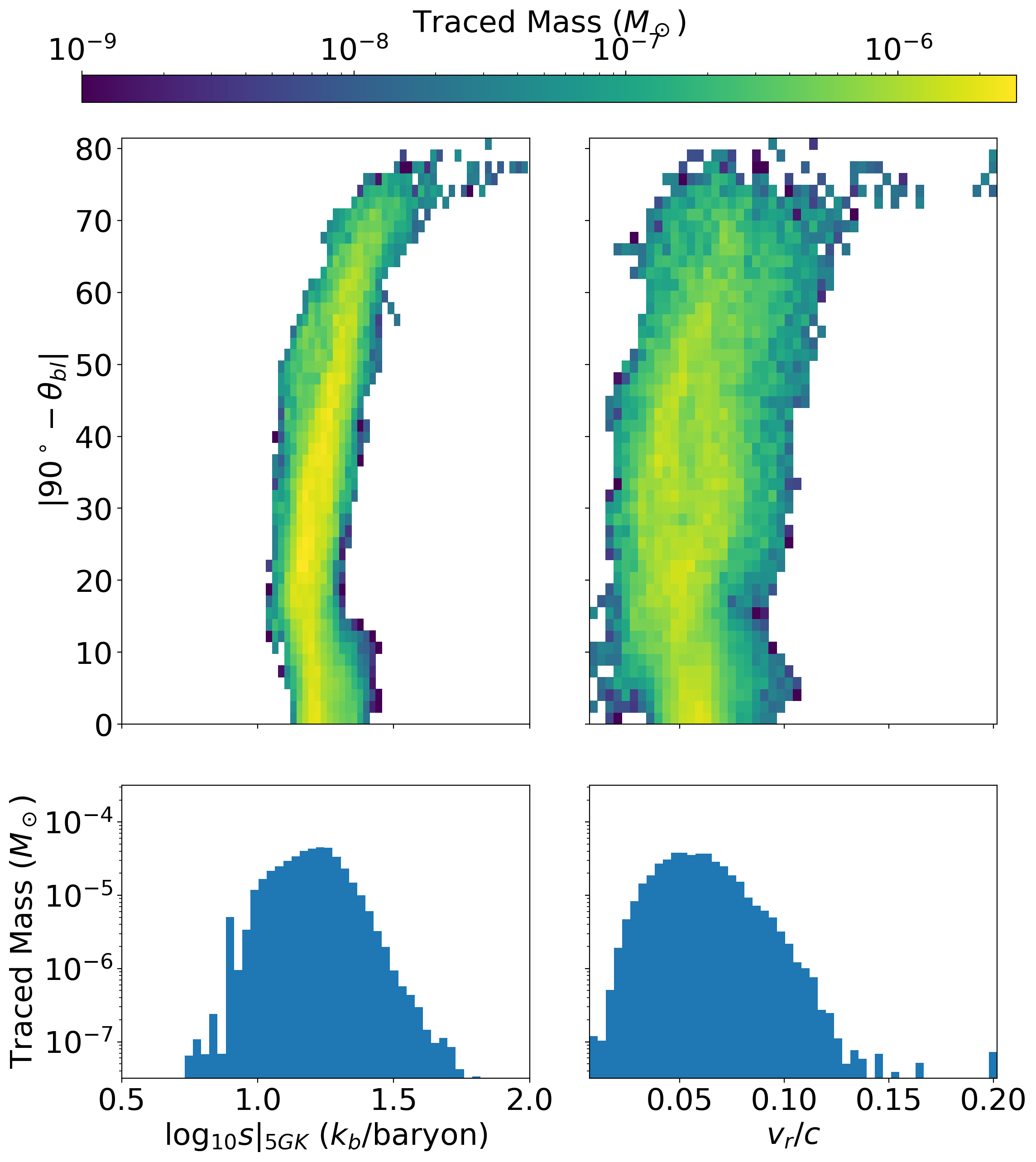}
  \caption{Histograms of entropy $s$ (left) and velocity (right) of
    gravitationally unbound material. The top row compares these
    quantities to distance from the mid-plane
    $|90^\circ - \theta_{bl}|$. The bottom simply bins by mass. Most
    of the material is slow moving and low entropy. However, there is
    a short tail to the distribution, with entropies as large as 100
    $k_b/$baryon and velocities as large as $0.2 c$. This tail is not
    statistically well-resolved by the number of tracer particles we
    use.}
  \label{fig:s:vr:v:theta}
\end{figure}

Figure \ref{fig:s:vr:v:theta} plots the entropy and radial velocity
$v_r=\partial r/\partial\tau$ (for proper time $\tau$) of
gravitationally unbound material, integrated over simulation time. The
angle and radial velocity are measured at an extraction radius of
$250 G M_{\rm BH}/c^2$ or about $1000$ km. At this radius, the average
tracer temperature is about $2GK$. The entropy is measured when the
material drops below a temperature of $5GK$. We find that most
material has low entropy, around 17 $k_b/$baryon, and a velocity of
about $0.05 c$. Both distributions have short tails, with entropies as
large as 65 $k_b/$baryon and velocities as large as $0.2c$. Note that
this is qualitatively different from the neutron star merger case,
where both disk wind and dynamical ejecta can move at a significant
fraction of the speed of light \citep{MillerGW170817}. Understanding
this difference will be the focus of future work.

\begin{figure}[tbp]
  \centering
  \includegraphics[width=\linewidth]{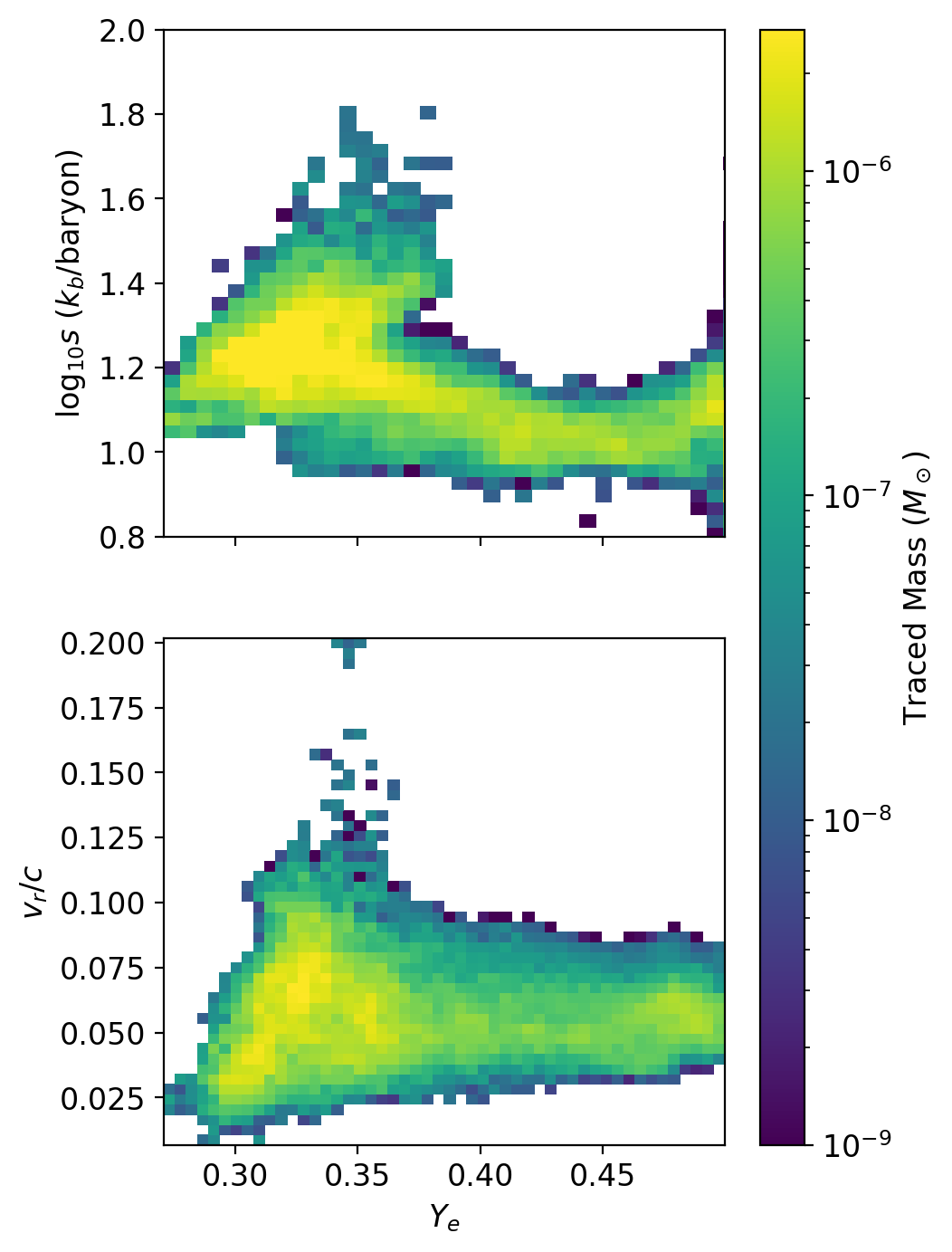}
  \caption{Histograms of entropy $s$ (top) and velocity (bottom)
    vs. electron fraction $Y_{\rm e}$ for gravitationally unbound material.}
  \label{fig:s:vr:v:ye}
\end{figure}

Figure \ref{fig:s:vr:v:ye} compares entropy and velocity with electron
fraction in the gravitationally unbound material. All material with
entropy greater than $\sim 20 k_b/$baryon has electron fraction of
$Y_{\rm e} \lessapprox 0.35$. Similarly, the minimum and maximum velocities
converge as $Y_{\rm e}$ increases, but the distribution is overall tighter
in velocity and very broad in $Y_{\rm e}$.

\subsection{Nucleosynthesis}
\label{sec:nuc}

For each of the gravitationally unbound tracer particles of
Sec.~\ref{sec:outflow}, we perform nucleosynthesis calculations using
the nuclear reaction network Portable Routines for Integrated
nucleoSynthesis Modeling (PRISM)~\citep{Mump2017,
  CoteRProcess,Cf254,Sprouse2019}.  For charged particle reaction
rates, we implement the Reaclib Database~\citep{REACLIB}. Neutron
capture rates are calculated using the Los Alamos National Laboratory
(LANL) statistical Hauser-Feshbach code of~\cite{Kawano2016}, assuming
nuclear masses of FRDM2012~\citep{FRDM2012}. Beta-decay properties are
similarly calculated using the LANL QRPA+HF
framework~\citep{MumpQRPA+HF, Mumpower2018,MOLLER20191}.  Finally, we
supplement these datasets with the nuclear decay properties of the
Nubase 2016 evaluation~\citep{NUBASE2016} and AME2016~\citep{AME2016}
where appropriate.

\begin{figure}[tbp]
  \centering
  \includegraphics[width=\linewidth]{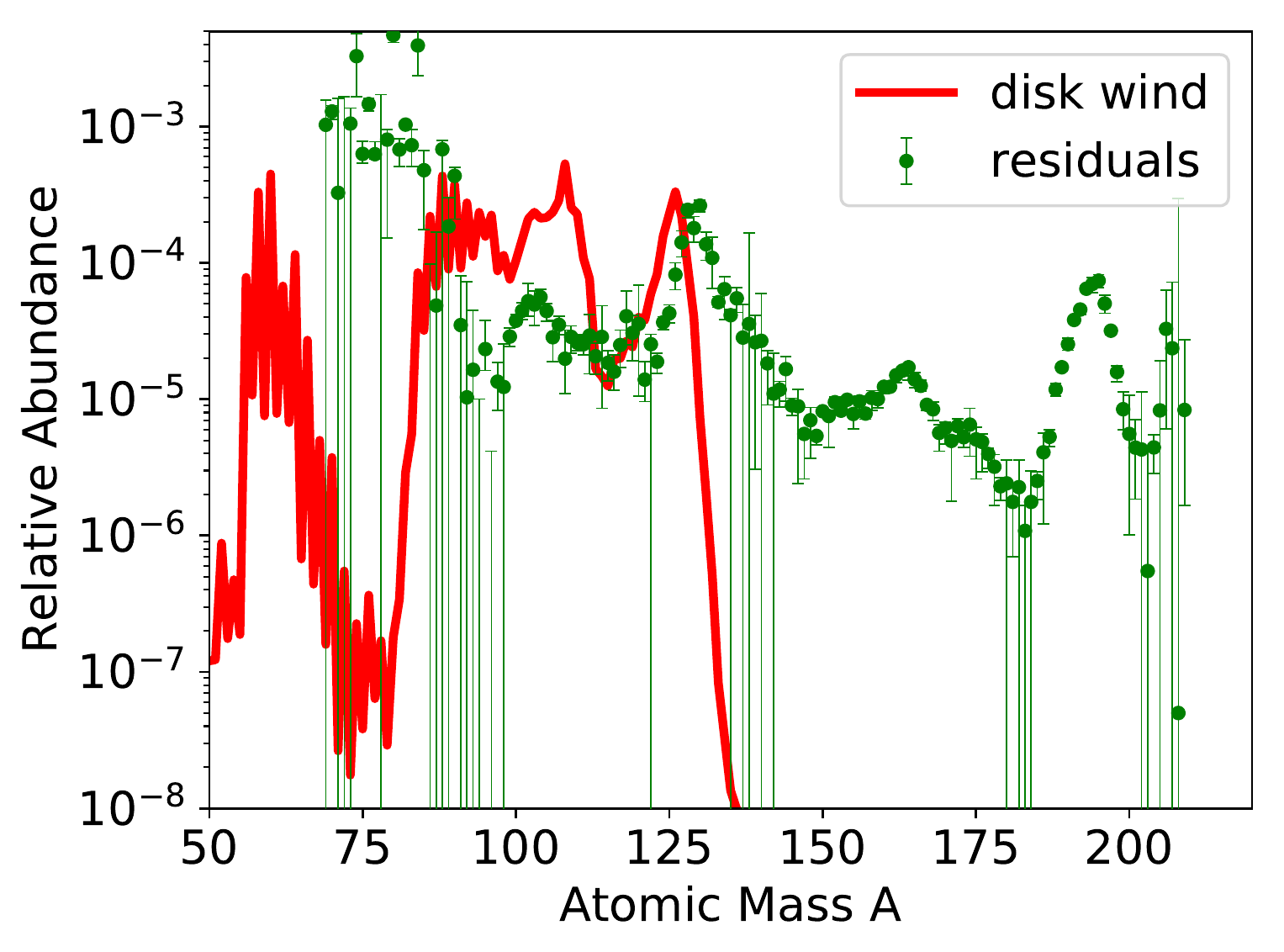}
  \caption{Mass-weighted abundances of r-process elements produced in
    outflow as a function of isotope mass A (red line). Green dots
    show r-process residuals measured from the solar system in
    \citet{Arnould+07}. The outflow produces first and second peak
    r-process elements, but no third-peak. Almost no elements with
    $A > 130$ are produced.}
  \label{fig:yeilds}
\end{figure}

Figure \ref{fig:yeilds} shows the mass-weighted nucleosynthetic yields
at $1$~Gyr.  As expected given the electron fraction and entropy
distributions, we find the outflow produces first- and (marginally)
second-peak elements but no third-peak elements. Indeed, almost no
elements with $A > 130$ are produced. Nucleosynthetic yields vary
greatly across individual tracer particles, however the overall
average abundances and the lack of heavy nuclei are robust.

\section{Systematics}
\label{sec:systematics}

Following \citet{SiegelCollapsar}, our model assumes that a phase of
fallback and subsequent accretion in a collapsar can be mapped to an
accretion disk that has relaxed from a compact torus with $Y_{\rm e}=0.5$ in
hydrostatic equilibrium \citep{FishboneMoncrief}. \footnote{Note that
  this torus is in \textit{hydrostatic} equilibrium. It is
  \textit{not} in equilibrium with neutrino radiation field.} (See
section \ref{sec:methods} for more details.) Here we examine this
assumption.

\subsection{How is Material Deleptonized?}
\label{sec:analytic-picture}

Material in our simulation deleptonizes close to the black hole as the
initial torus disrupts. In a real collapsar, material deleptonizes
as it falls back onto the black hole, potentially from far away. To
better understand the effects of the compact torus, we briefly
investigate models that do make a more
direct fallback assumption.

We examine the model of \citet{PophamNDAF} and \citet{dimatteo02},
which analytically incorporates several important neutrino emission
and absorption processes and a 5-piece equation of state. These models
assume a self-similar thin disk solution beginning at roughly 100km,
which forms the outer boundary condition of a steady-state $\alpha$
model \citep{ShakuraSunyaevAlpha}. Although this model neglects much
of the physics included in our simulation, the important difference in
boundary conditions makes it worth discussing.

\begin{figure}[tb]
  \centering
  \includegraphics[width=\linewidth]{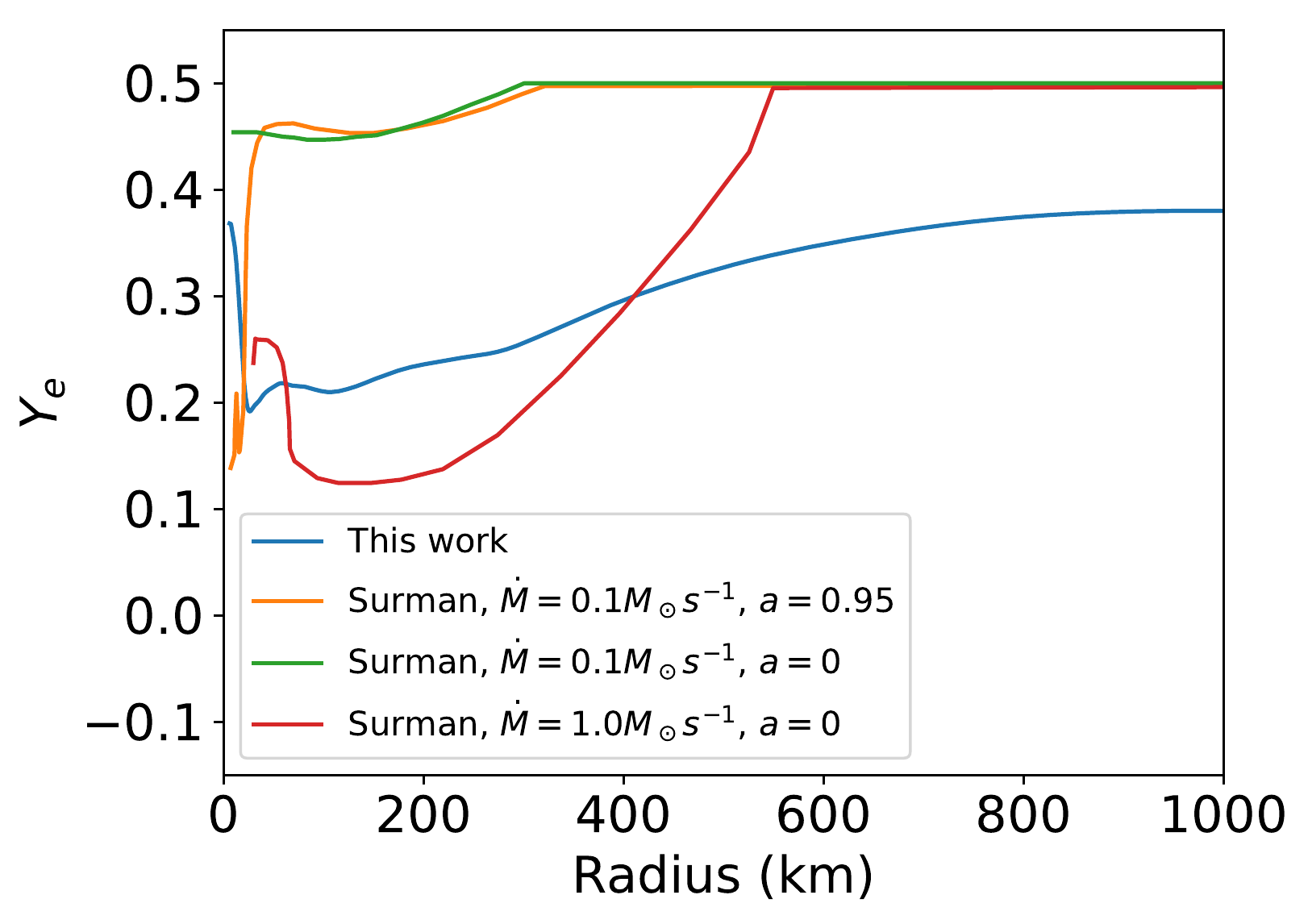}
  \caption{Electron fraction as a function of radius for several
    semi-analytic models, reproduced from \citep{surman04}. For
    comparison, we include the spherically averaged electron fraction
    computed in this work.}
  \label{fig:semi:analytic:Ye}
\end{figure}

Neutrino emission and absorption rates can be estimated from the
temperature and density in the disk. \citet{surman04} calculated the
electron fraction of the disk at a given radius by balancing the time
scale of deleptonization due to electron capture and subsequent
neutrino emission against the time required to accrete to a given
radius. Figure \ref{fig:semi:analytic:Ye} shows the electron fraction
for several analytic models computed by \citet{surman04}. For
comparison, we include the spherically averaged, density weighted
electron fraction
\begin{equation}
  \label{eq:def:Ye:SADW}
  \left\langle Y_{\rm e} \right\rangle_{\text{SADW}}(r) = \frac{1}{t_f - t_i}\int_{t_i}^{t_f} dt \frac{\int_{S^2} Y_{\rm e} \rho \sqrt{-g} d\Omega}{\int_{S^2} \rho \sqrt{-g}d\Omega}
\end{equation}
averaged from $t_i=5000 G M_{\rm BH}/c^3\approx 73$ ms to
$t_f = 10^4 G M_{\rm BH}/c^3\approx 147$ ms from this work, where $g$ is
the determinant of the metric and the integrals are over the
2-sphere. Recall that in the case of our full three-dimensional model,
we used a black hole of mass $3 M_\odot$ and a dimensionless spin
parameter of $a=0.8$. Note that these averages do not reflect the
diversity of physical conditions present in a simulation. They capture
the location of the disk as far as it has viscously spread, but they
do not show properties of, e.g., disk turbulence, the jet, or the
wind. Moreover, while the work of \citet{surman04} assumes a
time-independent solution, our system is of course time-dependent and
time-averaging does a poor job of capturing this. Figure
\ref{fig:time:evolution} shows the substantial variation in $Y_{\rm e}$
present in the simulation as a function of space and time.

Except in the innermost parts of the disk, the electron fraction of
the semi-analytic disk is well above the weak equilibrium value. This
is because the neutrino emission is very sensitive to the temperature
and is thus simply too slow to deleptonize the disk before it
accretes.

In the semi-analytic models, the in-falling matter is symmetric until
densities and temperatures rise sufficiently, at which point electron
fraction drops quickly. In contrast, we begin with a compact
torus. Ideally, after sufficient time, this disk achieves a
quasistationary state and ``forgets'' its initial conditions. However,
this assumption is incompatible with the inflow-deleptonization time
scale assumption in the semi-analytic models. Material in the disk has
\textit{already} deleptonized before viscously spreading outward as
it reaches a steady-state. Reconciling these two pictures requires
performing a full physics simulation with the correct in-fall initial
and boundary conditions.

\citet{mclaughlin05} and \citet{surman06} combined the electron
fraction in the disks of \citet{surman04} with a wind model and
calculated the composition of the wind ejecta, including the effects
of absorption. They found that as material leaves the disk, $Y_{\rm e}$
rises and that subsequently very little low $Y_{\rm e}$ material is ejected
in their models unless the accretion rate exceeds $1\,M_\odot
s^{-1}$. Their analysis also relies on a balance of fluid and weak
time scales, consistent with our 1D model \eqref{eq:1D:model}
calibrated to tracer data. Although these semi-analytic models are
very different from our full-physics simulation, the key result is
consistent.

\subsection{The Importance of Absorption for Nucleosynthesis }
\label{sec:absorption:nuc}

Although the core of the disk is low electron fraction, as described
in sections \ref{sec:outflow} and \ref{sec:nuc},  we find that
almost none of the gravitationally unbound material is sufficiently
low $Y_{\rm e}$ and high entropy to produce 3rd-peak r-process
elements.

\begin{figure}[tbp]
  \centering
  \includegraphics[width=\linewidth]{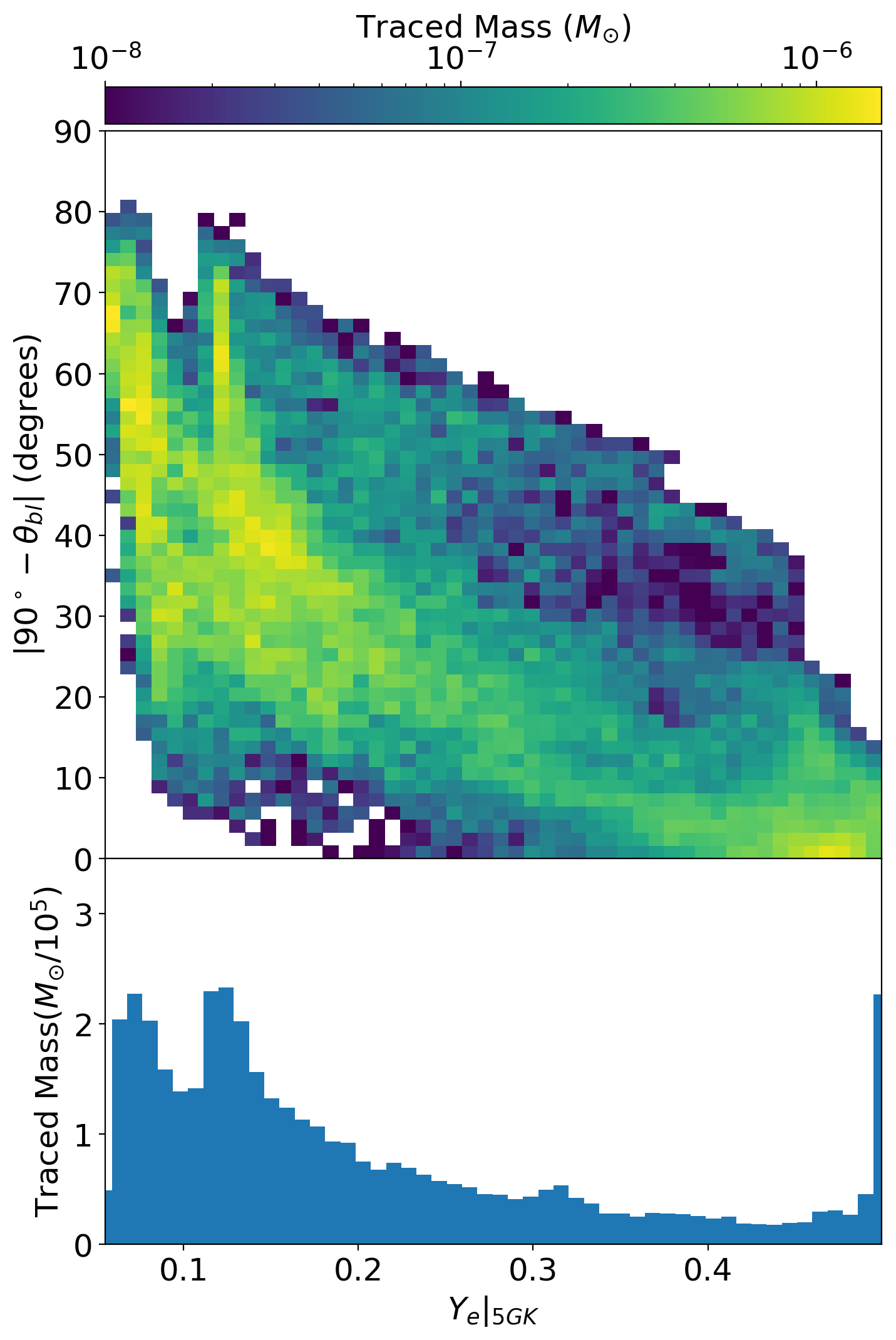}
  \caption{Electron fraction $Y_{\rm e}$ in the outflow, neglecting
    absorption. Top figure compares $Y_{\rm e}$ vs. angle and the bottom
    simply bins $Y_{\rm e}$, weighted by tracer particle mass. In contrast
    to the full transport case, if absorption is neglected, the
    outflow contains low electron fraction material. The spike in
    $Y_{\rm e}\approx 0.5$ is from viscous spreading at the back of the
    disk, which never drops to low electron fraction.}
  \label{fig:ye:em:hist}
\end{figure}

Figure \ref{fig:ye:em:hist} shows how this scenario changes if
absorption is neglected, which we calculate by separately tracking
changes in $Y_{\rm e}$ due to emission and absorption in our
simulation.\footnote{This calculation is performed in-line in
  \nubhlight. We separately record $Y_{\rm e}$ including both emission and
  absorption and $Y_{\rm e}$ with absorption neglected. The tracers record
  both quantities.} In the no-absorption scenario, the average
electron fraction in the outflow erroneously drops from
$\bar{Y}_e\approx 0.36$ to $\bar{Y}_e\approx 0.22$. This comparison
implies that including absorption in these models is critical to
correctly predicting nucleosynthetic yields in the outflow.

At late times, the neutrino optical depth is small, and yet we have
found that treating absorption is critical to capturing the electron
fraction in the outflow. As we showed in section
\ref{sec:absorption:early:time}, the physics of the early-time disk
depends on the interplay between neutrino emission and
absorption. Once the disk reaches a quasi-stationary state, optical
depths in the disk are low---of order $10^{-3}$. However, at early
times, during phases \textbf{(a)} and \textbf{(b)}, optical depths are
of order unity.

In other words, neutrino absorption in this early phase \textit{sets}
the initial conditions of the evolution in the quasi-stationary phase
\textbf{(c)} and thus strongly influences the electron fraction of the
outflow in the steady state. For the disk to reach the
\textit{correct} steady state, absorption opacity must be accounted
for.

\subsection{Effect of Stellar Envelope}
\label{sec:mass:escapes}

Section \ref{sec:outflow} assumes that all material with radius
$r_e > 250 G M_{\rm BH}/c^2$ and positive Bernoulli parameter is
gravitationally unbound. For an accretion disk in vacuum around a
black hole, this is likely a reasonable assumption. However, in a
collapsar scenario, the black hole-disk system is embedded in a
collapsing star. For the wind to escape, it needs not only to escape
the gravitational pull of the central black hole, but also of the star
itself. Also unaccounted for is the ram pressure of in-falling
material, as well as disruption of said fallback by the jet and
convection and advection from the dynamics of the fallback material.

These effects, alone or together, may significantly change the amount
of nucleosynthetic material which can escape the star. Understanding
this requires better modeling of the disk-wind-envelope system and
will be the subject of future work.

\section{Outlook and Implications}
\label{sec:conclusions}

We perform a three-dimensional general relativistic radiation
magnetohydrodynamics disk simulation of a nucleosynthetically
optimistic, high-accretion rate collapsar disk---the first to
incorporate full neutrino transport.

We find that a steady state disk forms with electron fractions near
the mid-plane as low as $Y_{\rm e}\sim 0.15$. At higher latitudes, however,
the electron fraction is significantly larger. This quasi-steady
accretion flow drives a relatively neutron-poor outflow; there is
almost no unbound material produced with electron fraction less than
$Y_{\rm e} \sim 0.3$.

We explore the evolution of the electron fraction of the disk from the
initial conditions through ``late''-time quasi-stationary flow. We
find the electron fraction rapidly drops as an angle dependence
appears in $Y_{\rm e}$. Over time, the electron fraction rises and
homogenizes as the accretion rate drops. This drop is an artifact of
the initial conditions and the fact that there is a finite reservoir
of material to accrete.

We present a simple one-dimensional model explaining the dependence of
the electron fraction within the flow on latitude and show that, when
properly calibrated with physically meaningful parameters, the model
matches the observed flow state extremely well. Although our analysis
is for collapsars, it may have relevance to disks formed after neutron
star mergers as well. We plan to pursue this avenue in future work.

We simulate r-process nucleosynthesis in this outflow via the PRISM
reaction network ~\citep{Mump2017,Cf254,Sprouse2019} and find almost
no material with an atomic mass above $A\sim 130$ is produced. Our
results thus imply that, even in the most nucleosynthetically
optimistic case, wind-driven off of accretion disks in collapsars
likely cannot act as a source for third-peak r-process
elements. Indeed, since collapsars produce first- and second-peak
r-process elements but not third-peak ones, including them as a
significant source of light r-process elements at all may be in
tension with the galactic chemical evolution and the solar abundance
pattern \citep{CoteRProcess}.

We compare our model to the GRMHD model of \citet{SiegelCollapsar} and
the semi-analytic models developed in the literature
\citep{PophamNDAF,dimatteo02,surman04,mclaughlin05}. We find the
electron fraction in our disk is significantly higher than reported in
\citet{SiegelCollapsar}, but lower than predicted in
\citet{surman04,mclaughlin05}. Moreover, we find that the electron fraction in our
outflow is consistent with the semi-analytic picture.

A potential confounding factor in understanding the electron fraction
in both the disk and the outflow is our use of a compact torus initial
condition. This torus construction is a standard in the disk
community. However, it is likely not appropriate for modeling a
collapsar. First, the compact torus chosen provides a reservoir of
material too close to the black hole, which means the gas does not
have time to naturally deleptonize as it accretes. Second, the compact
torus initial data ignores the presence of a star around and feeding
the disk. If the disk is fed, the accretion rate will not drop as the
power law found in this work. Rather it will depend on the mass
fallback rate. Moreover, as we discussed in section
\ref{sec:mass:escapes}, the stellar envelope may have a significant
effect on the mass in the outflow.

We argue that our discrepancy with \citet{SiegelCollapsar} is related
to how the initial conditions proceed to equilibrium when absorption
is or is not included. Including absorption allows us to more closely
match the flow state of a collapsar, where the disk is fed by
fallback.

However, an obvious improvement is to use an initial condition that
reflects the reality of a collapsing star. This strategy, adopted in
the early work of \citet{MacFadyen_1999}, would also allow us to
better understand exactly how much ejecta escapes the star. As
discussed in section \ref{sec:mass:escapes}, this is difficult to
address in a simulation that begins with an equilibrium torus. We will
pursue such a program in future work.

We conclude by emphasizing three takeaway messages from our
work. First, accurate treatments of neutrino transport and neutrino
absorption are required to capture the evolution of $Y_{\rm e}$. Second,
initial conditions should be carefully considered for collapsar
modeling. Finally, our model supports previous conclusions that even
under optimistic assumptions, wind blown off of accretion disks in
collapsars cannot act as a robust source of r-process material.

\section{Acknowledgements}
\label{sec:ack}

We thank Adam Burrows, Eliot Quataert, Charles Gammie, Jim Stone,
Philipp Moesta, Luke Roberts, Benoit Cote, Sam Jones, Wes Even, Oleg
Korobkin, Ryan Wollaeger, Sanjana Curtis, Greg Salvesen, and Daniel
Siegel for many helpful discussions. We also thank our anonymous
reviewer for challenging us to pursue a deeper analysis of the
dynamics of the disk.

We acknowledge support from the U.S. Department of Energy Office of
Science and the Office of Advanced Scientific Computing Research via
the Scientific Discovery through Advanced Computing (SciDAC4) program
and Grant DE-SC0018297

This work was supported by the US Department of Energy through the Los
Alamos National Laboratory. Additional funding was provided by the
Laboratory Directed Research and Development Program, the Center for
Space and Earth Science (CSES), and the Center for Nonlinear Studies
at Los Alamos National Laboratory under project numbers 20190021DR,
20180475DR (TS), and 20170508DR. This research used resources provided
by the Los Alamos National Laboratory Institutional Computing
Program. Los Alamos National Laboratory is operated by Triad National
Security, LLC, for the National Nuclear Security Administration of
U.S. Department of Energy under Contract No. 89233218CNA000001.

This article is cleared for unlimited release, LA-UR-19-30392.

We are grateful to the countless developers contributing to open
source projects on which we relied in this work, including Python
\citep{rossumPythonWhitePaper}, the GNU compiler
\citep{stallman2009using}, numpy and scipy \citep{numpy,scipyLib},
Matplotlib \citep{hunterMatplotlib}, the GNU scientific library
\citep{galassi2009gnu}, and HDF5 \citep{hdf5}.

% \appendix

%bibliography
% \newpage
\bibliography{nubhlight}
\bibliographystyle{apj}

\listofchanges

\end{document}